%% file: GammaICM.tex
\documentclass[final,5p,times,twocolumn,authoryear]{elsarticle}

\usepackage{epsfig}
\usepackage{verbatim}
\usepackage{epstopdf}
\usepackage{graphicx}
\usepackage{aas_macros}
\usepackage{epsf}
\usepackage{amsmath}
\usepackage{xcolor}
\usepackage{tikz}
\usepackage{longtable}
\usepackage{deluxetable}
\usepackage{threeparttable}
\usepackage{dblfloatfix}
\usepackage{enumitem}

\definecolor{darkgreen}{rgb}{0.0,0.5,0.0}
\usepackage[colorlinks,citecolor=darkgreen]{hyperref}

\input{Definitions.tex}

\newcommand{\Myfr}{{j}}
\newcommand{\Myc}{{\mathsf{c}}}
\newcommand{\MyC}{{\mathsf{C}}}
\newcommand{\Myh}{{\mathsf{h}}}
\newcommand{\MyH}{{\mathsf{H}}}
\newcommand{\mys}{{s}}
\newcommand{\Myn}{{\mathsf{n}}}
\newcommand{\Myf}{{\mathsf{f}}}
\newcommand{\myP}{{\mathbb{P}}}
\newcommand{\NS}{\mathcal{N}}
\newcommand{\FS}{\mathcal{F}}
\newcommand{\Mach}{\mathcal{M}}
\newcommand{\bmt}{{\boldsymbol{\theta}}}
\newcommand{\bmr}{{\boldsymbol{r}}}
\newcommand{\bmp}{{\boldsymbol{\psi}}}

\defcitealias{ReissKeshet18}{Paper I}
\newcommand{\PapI}{{\citetalias{ReissKeshet18}}}
\newcommand{\PapIS}{{\hyperlink{cite.ReissKeshet18}{I}}}

\defcitealias{Keshet25PaperII}{Paper II}
\newcommand{\PapII}{{\citetalias{Keshet25PaperII}}}
\newcommand{\PapIIS}{{\hyperlink{cite.Keshet25PaperII}{II}}}

\makeatletter
\def\ps@pprintTitle{
  \let\@oddhead\@empty
  \let\@evenhead\@empty
  \def\@oddfoot{\reset@font\hfil\thepage\hfil}
  \let\@evenfoot\@oddfoot
}
\makeatother

\newcommand{\SetCurrentFigure}[1]{\def\currentfig{#1}}
\newcounter{panel}

\newcommand{\panellabel}[1]{\refstepcounter{panel}\label{#1}}

\begin{document}

\begin{frontmatter}

\title{Galaxy-cluster-stacked Fermi-LAT III: substructure and radio-relic counterparts}

\author{Uri Keshet}
\address{
    Physics Department, Ben-Gurion University of the Negev, POB 653, Be'er-Sheva 84105, Israel; keshet.uri@gmail.com
}

\begin{abstract}
Faint $\gamma$-ray signatures emerge in \emph{Fermi}-LAT data stacked scaled to the characteristic $R_{500}$ radii of MCXC galaxy clusters.
This third paper in a series shows a $4.3\sigma$ excess of discrete 4FGL-DR4 catalog $\gamma$-ray sources at the $r<1.5R_{500}$ radii of 205 clusters,
coincident with an $r\sim R_{500}$ diffuse $2.6\sigma$ excess of 1--100 GeV emission from 75 high-latitude clusters.
The source excess becomes highly ($>5\sigma$) significant when considering the substantial ($3.4\sigma$) and unexpectedly rapid quenching of $\gamma$-ray sources just inside the virial shock.
The excess sources show radial, spectral, and luminosity distributions better matching radio-relic counterparts or substructure than present tentative classifications as blazar-candidates.
Their spectral distribution is bimodal: flat-spectrum sources are consistent with enhanced hadronic emission behind weak, Mach $\sim2$ shocks, while softer sources may be phoenix counterparts.
\end{abstract}

\end{frontmatter}

\section{Introduction}

Cosmic-ray (CR) ions (CRI) cool slowly, so they accumulate in the intracluster medium (ICM) of groups and clusters of galaxies (henceforth clusters).
Hadronic models for radio minihalos \citep{PfrommerEnsslin04, KeshetLoeb10}, giant halos \citep{Dennison80, BlasiColafrancesco99, KushnirEtAl09}, and standard relics \citep{Keshet10} require similar CRI densities and spectra in different environments and dynamical states.
This observation, supported by additional evidence, suggests that clusters harbor a spatially and spectrally flat, $du/d\ln E\simeq \const.$, CRI population, which powers the diffuse ICM radio emission in its various forms \citep{Keshet10}, including transitional states (minihalo--giant halo growth and halo--relic bridges), and the recently-discovered mega-halos \citep{Keshet24}, and implies a range of nonthermal X-ray and \gama-ray counterparts.
Here, $E$ and $u$ are the CRI energy and energy density.

In this joint hadronic model, synchrotron emission by secondary CR electrons (CRE) from the decay of charged pions produced in the inevitable inelastic $p$-$p$ collisions of CRI with ambient nuclei, traces in radio all ICM regions where magnetization is sufficiently strong.
In particular, hadronic standard radio relics (shock-driven type; also known as gischt/flotam/radio shocks; henceforth shock relics) are expected behind weak ICM shocks, explaining why their radio spectrum is inconsistent with primary CRE acceleration in a weak shock \citep{Keshet10}.
CRI delayed in the near downstream before rapidly diffusing away then account for the dimming and spectral softening found downstream of such relics.
Indeed, the required strong CR diffusion, measured in halos and shock-relic downstreams, also explains the inferred homogeneity of the CRI and the very soft spectra observed in the periphery and high-frequency tails of halos \citep{Keshet24}.

The joint hadronic model predicts that shock relics, and not only halos, should have non-thermal, spectrally-flat counterparts at high photon energies, in X-rays from Compton emission and in \gama-rays mainly from $\pi^0\to\gamma\gamma$ decay.
Phoenix-type radio relics, also known as shocked active-galactic-nuclei (AGN)/fossils/ghosts (henceforth phoenix relics), can also be explained as hadronic emission, but from fossil CRI; this model fits the data better than leptonic alternatives \citep{Keshet25Phoenix}.
The hadronic phoenix-relic model predicts nonthermal X-rays and \gama-rays similar to shock relics, but softer, and detectable mainly in the harder-spectrum phoenixes.

It should be noted that hadronic radio models in general have been increasingly dismissed in recent years or even decades, in favor or leptonic models that invoke electron (re)acceleration in turbulence or weak shocks, due to spectral arguments \citep[refuted in][]{Keshet24} and upper limits on cluster \gama-rays; for a review, see \citet{vanWeerenEtAl19}.
Nevertheless, detections of central \gama-ray emission from the Coma cluster \citep{XiEtAl18_Coma, AdamEtAl21, BaghmanyanEtAl22} are consistent with the joint hadronic model when a flat spectrospatial CRI distribution is considered \citep[][]{Keshet24, KushnirEtAl24}.

Moreover, Paper II \citep{Keshet25PaperII} of this series, which is devoted to amplifying cluster \gama-ray signals by stacking \emph{Fermi}-LAT data over clusters scaled to their characteristic $R_{500}$ radii, recently found a significant, central, extended \gama-ray signal at dimensionless $\tau\equiv r/R_{500}\lesssim 0.5$ radii. 
Here, $R_{500}$ encloses a mean mass density $500$ times the critical density of the Universe.
This signal, emerging after AGN and discrete sources were carefully removed, is best fit by an approximately constant $du/d\ln E=10^{-13.6\pm0.5}$ erg cm$^{-3}$, matching the joint hadronic-model predictions \citep{Keshet10} in terms of spectrum, spatial distribution, and normalization.

The detection of this subtle signal, validating the joint hadronic model over a wide range of clusters, was facilitated by the approximate similarity of clusters when scaled by a uniformly-estimated characteristic radius, such as $R_{500}$ in the Meta-Catalog of X-ray detected Clusters of galaxies \cite[MCXC;][]{PiffarettiEtAl11}.
Such extended or cluster scale-dependent signals weaken when stacked without $R_{500}$ scaling or when the effective resolution is degraded, and tend to soften if poorly-resolved clusters are included \citep{Keshet25PaperII}.

\begin{figure}
    \centering
    \includegraphics[width=0.45\textwidth]{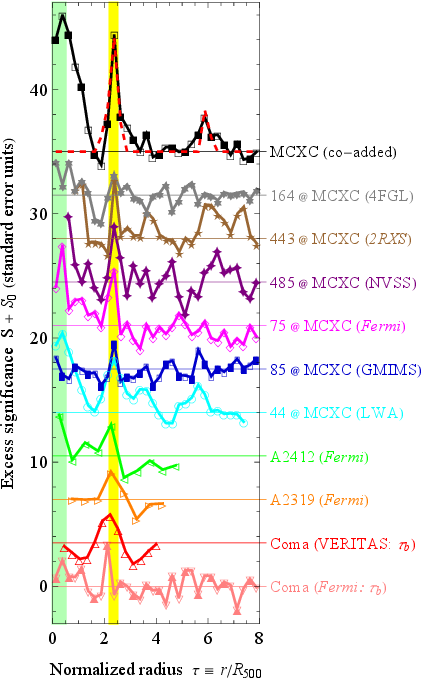}
	\caption{ \label{fig:VSummary}
        The \emph{Fermi}-LAT signal of {\PapII} (stacked over 75 clusters; magenta diamonds), analyzed here in the intermediate region between central CRI ($\tau<0.5$; vertical green band) and virial-shock CRE ($2.2<\tau<2.5$; yellow band; {\PapI}) signals, along with the distribution of 4FGL-DR4 discrete \gama-ray sources (six-pointed gray stars).
        Each signal is shown as the significance $S$ (symbols with lines to guide the eye; standard-error units) of the excess above the background $S_0$ (labelled horizontal lines, shifted vertically for visibility).
        Similar signals are found when stacking (bottom to top labels specify sample sizes) diffuse emission (empty symbols), discrete sources (filled), or without separating the two (intermittent empty and filled symbols): in LWA \cite[circles;][]{HouEtAl23} and polarized GMIMS \cite[rectangles;][]{Keshet24GMIMS}, and in NVSS (four-pointed stars) and 2XRS (five-stars) source catalogs \citep{IlaniEtAl24a}; the co-added MCXC excess (black squares) is consistent with a stacked cylindrical shock model \cite[dashed red curve;][]{Keshet24GMIMS}.
        Also presented (triangles) are individual clusters (bottom to top) Coma, in LAT \cite[down triangles;][]{keshet2018evidence} and VERITAS \cite[virial excess coincident with SZ drop and synchrotron; up triangles;][]{KeshetEtAl17} data as a function of $\tau_b$, and A2319 (right triangles) and A2142 (left triangles) in LAT data as a function of $\tau$ \citep{keshet20coincident}, all showing a virial excess coincident with a radial SZ drop.
	}
\end{figure}

The sensitivity of this method was demonstrated earlier, in Paper I \citep{ReissKeshet18} of this series, which detected cluster virial shocks in a narrow, $2.2<\tau<2.5$ radial range through Compton emission from their accelerated CRE.
Followup studies, demonstrated in Fig.~\ref{fig:VSummary}, verified that virial shocks give rise to a noticeable signal at these radii in multiple data sets and in different anticipated channels, including Compton emission, polarized and non-polarized synchrotron, and the Sunyaev-Zel'dovich (SZ) effect, providing new constraints on such non-magnetized collisionless shocks \citep{KeshetHou24}.
Surprisingly, simply stacking cataloged discrete, radio or X-ray sources around scaled cataloged clusters also shows (star symbols in the figure) a strong coincident excess, probably from galactic halos and galactic outflows energized by the shock \citep{IlaniEtAl24a, IlaniEtAl24}.

In this Paper III of the series, we stack both cataloged \gama-ray sources and diffuse LAT data, around MCXC clusters scaled to their $R_{500}$ radii, in order to address interrelated questions motivated by the preceding introduction:
\begin{enumerate}[leftmargin=*, itemsep=0pt]
\item
Does stacking discrete sources from the most recent \gama-ray catalog, namely the \emph{Fermi}-LAT Fourth Source Catalog \citep[4FGL, Data Release 4;][]{FGL4DR4}, around scaled clusters, show the virial shock and other signals of interest?
\item
As {\PapII} identified the \gama-ray counterparts of minihalos and giant halos, can \emph{Fermi}-LAT stacking also detect the \gama-ray counterparts of radio relics, of both shock and phoenix types, arising from enhanced pion production?
\item
Is ICM substructure, such as dense clumps and CR-rich bubbles, detectable by \emph{Fermi}-LAT, given the detection of the $\sim$homogeneous CRI distribution permeating the ICM?
\item
What is the nature of the $2.6\sigma$ excess in diffuse \emph{Fermi}-LAT emission, found in Papers {\PapIS} and {\PapIIS} at intermediate, $\tau\simeq 1$ radii, between the central and virial-shock signals? (See elevated diamonds between the two vertical bands in Fig.~\ref{fig:VSummary}.)
\end{enumerate}

In \S\ref{sec:Methods}, we outline the data and methods used to stack and analyze both discrete 4FGL-DR4 sources and diffuse \emph{Fermi}-LAT data.
A weak excess of discrete sources towards the virial shock, and a strong quenching of such sources just inside the shock, are described in \S\ref{sec:Quenching}.  A strong excess of sources in the ICM, especially considering said quenching, is analyzed in \S\ref{sec:ICM}.
Both anticipated and measured properties of \gama-ray ICM sources are considered in \S\ref{sec:Props}.
Finally, the weak excess in diffuse emission from $r\sim R_{500}$ radii, coincident with a peak in discrete sources, is studied in \S\ref{sec:Diffuse}.
The results are summarized and discussed in \S\ref{sec:Discussion}.
\ref{app:Sensitivity} demonstrates the robustness of the results, and \ref{app:Prop} provides additional properties of the excess ICM sources, which are listed in \ref{app:Sample}.

As in previous papers in this series, the same flat $\Lambda$CDM cosmological model used in MCXC is adpoted, with a present-day Hubble constant $H_0=70\km\se^{-1}\Mpc^{-1}$ and a mass fraction $\Omega_m=0.3$; an $f_b=0.17$ baryon fraction is also assumed.
Confidence intervals are $68\%$ containment projected onto one parameter, unless otherwise stated (in \S\ref{sec:Diffuse}).

\section{Data and methods: source and photon stacking}
\label{sec:Methods}

We stack, bin, and analyze the discrete 4GFL-DR4 sources in the method of \citet{IlaniEtAl24a}, and the diffuse LAT data in the method of {\PapI}, recently updated in {\PapII}.
These methods are discussed and justified in detail in the above references, which also demonstrate their robustness, so here we provide only a brief overview.

Data are stacked around MCXC clusters of characteristic angles $0.2<\theta_{500}<0.5$, limited from below by the LAT point spread function \citep[PSF; \eg][]{AtwoodEtAl13} and from above by foreground structure and point-source abundance.
Here, $\theta_{500}\equiv R_{500}/d_A$, where $d_A$ is the angular diameter distance of the cluster.
Galactic foreground contamination is lowered by selecting clusters at $|b|>20\dgr$ Galactic latitudes; a smaller but cleaner sample of $|b|>30\dgr$ clusters is also considered.

Following the previous papers in this series, we define the $\tau<15$ disk around each cluster as its region of interest (ROI).
This choice of outer radius is approximately twice the maximal radius at which cluster-related signals can be seen, in particular the tentative, $3.1\sigma$ excess at $\tau\simeq 6$, modeled in Fig.~\ref{fig:VSummary} as the possible outer radii of elongated virial shocks.
Hence, a sufficiently large, $7<\tau<15$ region remains in the ROI for foreground and background (field, henceforth) estimation.
Similarly, $\Delta\tau=1/4$ is again adopted as the nominal radial bin size, although higher and lower resolutions are also examined.

\subsection{Stacking catalog sources}

Consider the 4FGL-DR4 sources found in the ROI of a sample $\MyC$ of $N_c$ clusters.
Denoting $\NS(\tau,\Myc)$ as the number of sources found in a projected radial ring of normalized radius $\tau$ and width $\Delta\tau$ around cluster $\Myc\in\MyC$, and $\FS(\tau,\Myc)$ as the number of field sources anticipated in this ring, we may adopt a strong field-only, $\NS(\tau,\Myc)=\FS(\tau,\Myc)$ null hypothesis.
Alternatively, summing $\NS(\tau)\equiv\sum_{\Myc=1}^{N_c}\NS(\tau,\Myc)$ over the sample clusters, we adopt a weaker, averaged, $\NS(\tau)=\FS(\tau)\equiv\sum_{\Myc=1}^{N_c}\FS(\tau,\Myc)$
null hypothesis.
For transparent small-number statistics, these $\NS$ and $\FS$ definitions avoid normalization by the $N_c$ sample size.

The dimensionless surface number density of observed sources, $\mathfrak{n}(\tau) \equiv d\NS/d\Omega'\simeq \NS/\Delta \Omega' \simeq \NS/(2\pi\,\tau\,\Delta \tau)$,
where $d\Omega'\equiv 2\pi\theta_{500}^{-2}\sin(\theta)d\theta$ is the dimensionless solid angle differential, approaches a constant $\mathfrak{f}$ far from the cluster, at $\tau\gtrsim 7$ radii, so for simplicity we adopt and uniform field model and approximate $\FS\simeq \mathfrak{f}\,d\Omega'$.
Figure \ref{fig:Excess1} shows the $\mathfrak{n}$ (circles) and $\mathfrak{f}$ (dashed line) surface densities, for a lower quality analysis (panel b), stacking all 4FGL-DR4 sources around the $|b|>20\dgr$ sample of $205$ clusters (of median redshift $z\simeq 0.037$ and mass $M_{500}\simeq 0.86\times 10^{14}M_\odot$ values), and for a higher quality (panel c) analysis, stacking only significant, $>5\sigma$ sources around the $|b|>30\dgr$ sample of $164$ clusters (of median $z=0.036$ and $M_{500}\simeq 0.88\times 10^{14}M_\odot$).

A positive, $\mathfrak{n}>\mathfrak{f}$ excess can be assigned a significance $S\simeq (\NS-\FS)/\FS^{1/2}$ (standard error units) in the $\FS\gg1$, normal-distribution limit.
However, at the small radii of interest, $\FS$ becomes small, so Poisson statistics are necessary.
As the probability of finding $\NS$ sources in a bin then becomes $e^{-\FS} \FS^\NS/\NS!$, we may associate an excess $\mathfrak{n}>\mathfrak{f}$ of confidence level $S>0$ (standard error units) with the surface density $\NS_S$ satisfying $1-\Gamma(\NS_S,\FS)/\Gamma(\NS_S)=p\mbox{-value}=\mbox{erfc}(S/2^{1/2})/2$, where $\mathrm{erfc}(z)$, $\Gamma(z)$, and $\Gamma(a,z)$ are respectively the complementary error, Euler gamma, and incomplete gamma functions.
Similarly, for an $\mathfrak{n}<\mathfrak{f}$ deficit of sources, we may associate an $S<0$ confidence level with the $\NS_S$ value satisfying
$\Gamma(1+\NS_S,\FS)/\Gamma(1+\NS_S)=p\mbox{-value}=\mbox{erfc}(-S/2^{1/2})/2$.

Figure \ref{fig:VSummary} shows (gray six-pointed stars) the $S(\tau)$ profile implied by the measured $\FS$ and $\NS(\tau)$ profile in our higher quality stacking, of significant 4FGL-DR4 sources around $164$ high-latitude clusters.
Figure \ref{fig:Excess1} shows (in panels b and c), in addition to $\mathfrak{n(\tau)}$ and $\mathfrak{f}$, also the $\pm1\sigma, \pm2\sigma, \pm3\sigma, \ldots$ surface density levels (dotted curves) implied by the measured $\FS$ in each panel.

\subsection{Stacking diffuse data}

For the diffuse stacking, we analyze updated archival, Pass-8 \citep[P8R3;][]{BruelEtAl18Pass8R3} LAT data from the Fermi Science Support Center (FSSC)\footnote[1]{\texttt{http://fermi.gsfc.nasa.gov/ssc}} using updated Fermi Science Tools (version \texttt{2.2.0}).
Pre-generated weekly all-sky files spanning weeks $9$--$422$ ($7.9\yr$) of the mission, during which the empirical PSF of {\PapII} retained good performance, are binned using the P8R3\_ULTRACLEANVETO\_V3 instrument response functions (IRFs) onto $N_\epsilon=4$ logarithmic energy bands in the $\epsilon=1\mbox{--}100\GeV$ photon energy range (labelled below as channels 1 through 4).
Sky maps are discretized using a HEALPix scheme \citep{GorskiEtAl05} of order $N_{hp}=10$, providing a $\delta\Omega\simeq 10^{-6}\sr$ pixel solid angle.
HEALPix pixels within the $95\%$ PSF containment of any of the 5697 significant ($>5\sigma$) 4FGL-DR4 sources are then masked, to minimize point-source contamination.

Following {\PapII}, a nominal sample of 75 high-latitude ($|b|>30\dgr$) clusters (median $z\simeq0.038$ and $M_{500}\simeq0.92\times 10^{14}$) is used to study the overall four-channel diffuse signal at different $\Delta\tau$ resolutions.
Modelling the more central, $\tau<1.5$ signals requires a high, $\Delta\tau=1/8$ resolution and a high-quality sample of 31 non-compact, $\theta_{500}>0\dgrdot25$ clusters (median $z\simeq0.028$ and $M_{500}\simeq0.86\times 10^{14}$), but then only channels $1$--$3$ have sufficient photon statistics.
Both samples exclude the two over-dense MCXC regions identified in {\PapI}: 12 clusters near Galactic coordinates $(l,b)=(315\dgr,32\dgr)$ and 4 clusters near $(12\dgr,50\dgr)$.
Also excluded, out of an abundance of caution, are all clusters with any $>5\sigma$ well-localized ($95\%$ containment semi-major axis $a_{95}<0\dgrdot08$) 4FGL-DR4 source within $1\dgrdot8$ (the $95\%$ containment angle at $1\GeV$) from the center of the cluster.
For these reasons, the cluster samples used for stacking diffuse data are smaller than their source-stacking counterparts.

For each channel $\Myfr$, cluster $\Myc$, and non-masked HEALPix pixel $\Myh$, we then define the excess $\Delta \Myn_{\Myfr}(\Myc,\Myh) = \Myn_{\Myfr}(\Myh)-\Myf_{\Myfr}(\Myc,\Myh)$ in the number $\Myn$ of detected photons with respect to the expected number $\Myf$ of field photons.
The mean excess brightness per cluster for a sample $\MyC$ of $N_c$ clusters then becomes
\begin{equation}
\label{eq:ExcessIj}
I_{\Myfr}(\tau; C; f)
= \frac{\epsilon_{\Myfr}\sum_{\Myc\in\MyC} \sum_{\Myh\in \MyH} \Delta \Myn_{\Myfr}(\Myc,\Myh)}{N_c \delta\Omega \sum_{\Myc\in\MyC} \sum_{\Myh\in \MyH} \mathcal{E}_{\Myfr}(\Myh) } \coma
\end{equation}
where $\epsilon_{\Myfr}$ is the average photon energy in channel $j$, $\MyH(\tau,\Myc)$ are the non-masked HEALPix pixels falling in the $\tau$ bin of cluster $\Myc$, and $\mathcal{E}_{\Myfr}(\Myh)$ is the exposure (\ie time-integrated effective area) of pixel $\Myh$ in channel $\Myfr$.

The field distribution $\Myf$ is evaluated around each cluster $\Myc$ by fitting the sky-distribution of the brightness $I_{\Myfr}(\{\tau_x,\tau_y\}; \{\Myc\};0)$ within the ROI as a polynomial of nominal order $N_f=2$ in the normalized sky coordinates $\{\tau_x, \tau_y\}$.
The choice of $N_f=2$ order, as well as other parameter values used in the analysis, do not strongly affect the results, as shown in Papers {\PapIS} and {\PapIIS}.

Figure \ref{panel:a} presents the resulting channel-averaged brightness profile,
\begin{equation}
\label{eq:ExcessI}
\langle I_{\Myfr}\rangle(\tau; C; \Myf)
= \frac{1}{N_\epsilon} \sum_{j=1}^{N_\epsilon} I_{\Myfr} \coma
\end{equation}
for low (circles; $\Delta\tau=1/2$) vs. high ($\Delta \tau=1/8$) resolutions, with nominal analysis parameters, stacked over the 75-cluster sample using all four channels.
The significance of the excess brightness corresponding to this profile,
derived in {\PapII}, is shown in Fig.~\ref{fig:VSummary} (magenta diamonds).

\section{Virial shock and quenching}
\label{sec:Quenching}

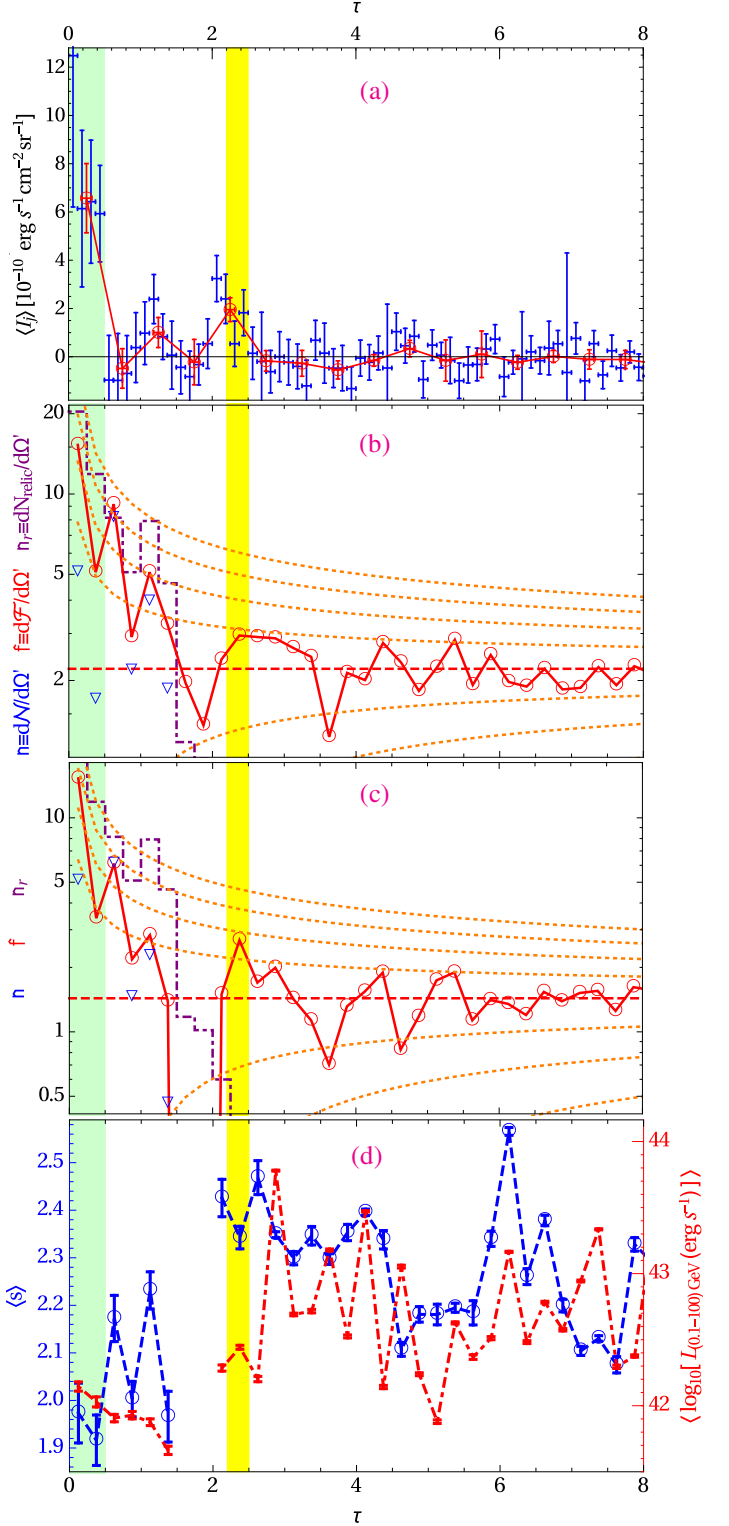
\begin{figure}
    \centering
        \vspace{-0.5cm}
        \SetCurrentFigure{2}
        \panellabel{panel:a}
        \begin{tikzpicture}
            \draw (0, 0) node[inner sep=0]
            {
                \hspace*{-0.85cm}
                \includegraphics[width=0.458\textwidth, trim={0 0.8cm 0 0}, clip]{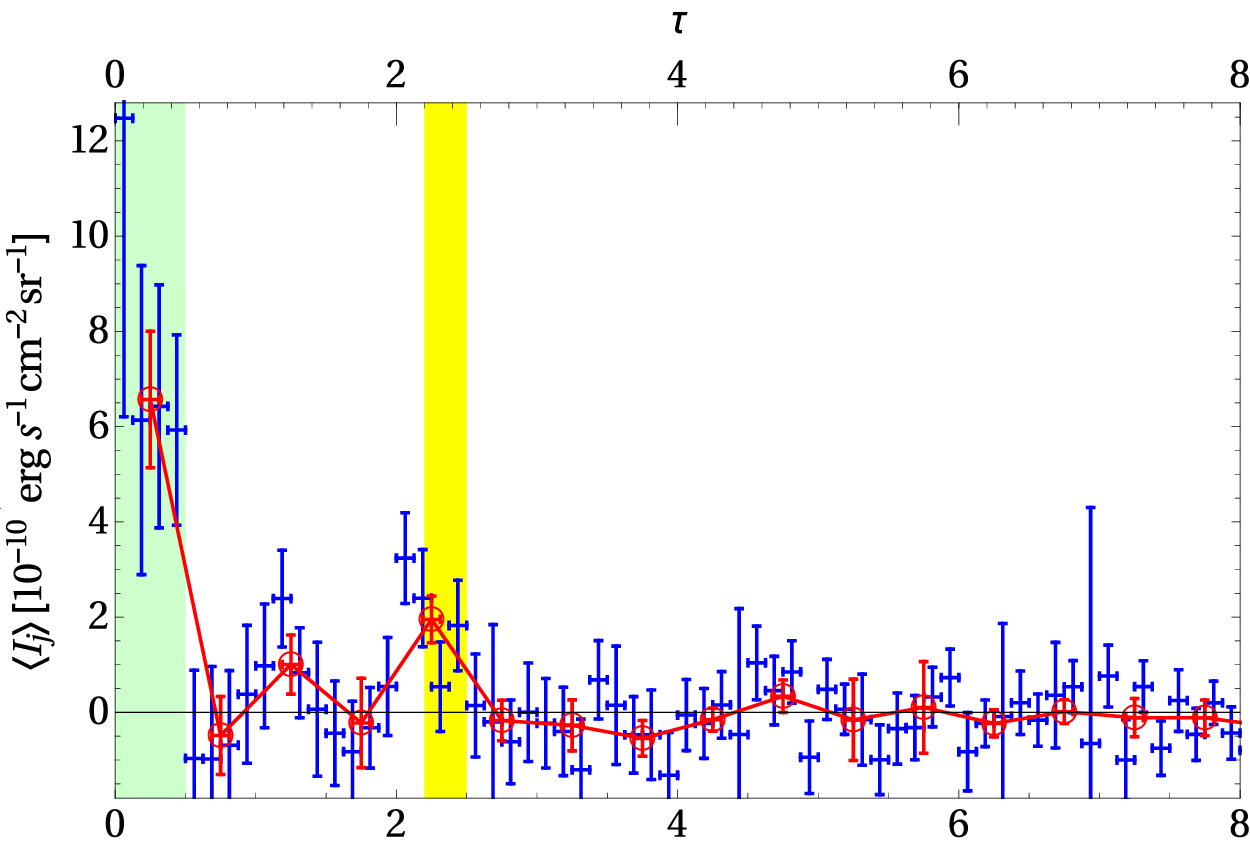}
            };
            \draw (0.22,1.4) node[text=magenta] {\normalsize (a)};
        \end{tikzpicture}
        \panellabel{panel:b}
        \begin{tikzpicture}
            \draw (0, 0) node[inner sep=0]
            {
                \hspace*{-0.8cm}
                \includegraphics[width=0.455\textwidth, trim={0 1.22cm 0 0}, clip]{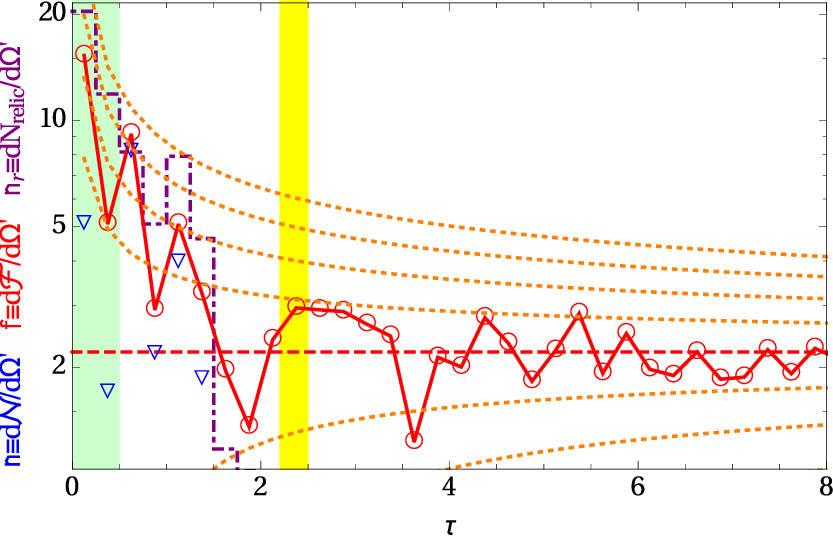}
            };
            \draw (0.22,1.8) node[text=magenta] {\normalsize (b)};
        \end{tikzpicture}
        \panellabel{panel:c}
        \begin{tikzpicture}
            \draw (0, 0) node[inner sep=0]
            {
                \hspace*{-0.87cm}
                \includegraphics[width=0.458\textwidth, trim={0 1.18cm 0 0}, clip]{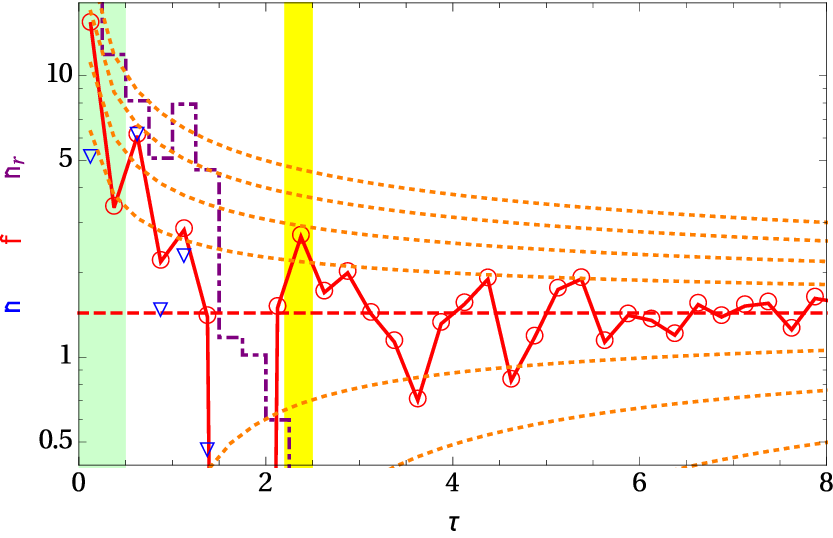}
            };
            \draw (0.22,1.9) node[text=magenta] {\normalsize (c)};
        \end{tikzpicture}
        \panellabel{panel:d}
        \begin{tikzpicture}
            \draw (0, 0) node[inner sep=0]
            {
                \hspace*{-0.28cm}
                \includegraphics[width=0.5\textwidth, trim={0 0 0 0.01cm}, clip]{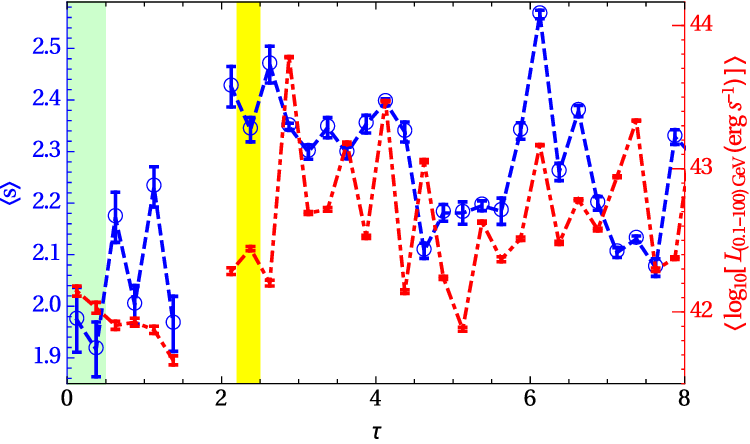}
            };
            \draw (0.05,2.2) node[text=magenta] {\normalsize (d)};
        \end{tikzpicture}
        \vspace{-0.8cm}
    \caption{
        Panel a: Diffuse channel-averaged brightness stacked at high ($\Delta\tau=1/8$; circled blue bars) vs. low ($\Delta \tau=1/2$; red bars)
        resolution.
        Panels b--d: stacked 4FGL-DR4 sources.
        Surface number densities, in nominal (panel b) and high-quality ($>5\sigma$ sources around $|b|>30\dgr$ clusters; panel c) stacking (red circles), vs. estimated field (dashed red, with $\pm1\sigma, \pm2\sigma,\ldots$ dotted confidence levels), sources without SIMBAD optical AGN associations (blue triangles), and shock relics (for $R_{500}=1\Mpc$; dot-dashed purple).
        Panel d: mean-weighted \gama-ray spectra $s$ (circled blue bars with left axis) and luminosities $L$ (red bars; right axis) for panel c sources.
        Vertical bands: as in Fig.~\ref{fig:VSummary}.
    \label{fig:Excess1}
	}
\end{figure}

As Fig.~\ref{fig:VSummary} shows, a narrow virial-shock excess at $\tau\simeq 2.4$, seen in multiple tracers, may correspond to a minimal virial shock radius, whereas a marginal broad excess at $\tau\simeq 6$ may reflect the more distant part of elongated shocks; a stacked cylindrical shock model \citep[dashed red in the figure; see][]{Keshet24GMIMS} fits both.
The significant \gama-ray brightness excess at $\tau\simeq 2.4$, seen in Figs.~\ref{fig:VSummary} and \ref{panel:a}, is not addressed here further, because it was already discussed in detail in {\PapI}, and recently updated in {\PapII}.
It is interesting, however, to note that this signal precisely coincides with a low-significance, $1.7\sigma$ excess of 4FGL-DR4 sources, seen in Fig.~\ref{fig:Excess1} (panels b and c).

While this $\tau\simeq 2.4$ source excess is of low significance, and there is no source excess at $\tau\simeq6$, Fig.~\ref{panel:d} shows that the sources suddenly become considerably softer, on average, around $\tau\simeq 6$, and less luminous around $\tau\simeq 2.5$.
This panel presents, for each radial bin, the (error-weighted) mean values of the \gama-ray spectral index $\mys\equiv 1-d\ln I_\epsilon/d\ln\epsilon$ and the luminosity $L=4\pi d_L^2 F$ in $100\MeV<\epsilon<100\GeV$ (received) photon energies (henceforth).
Here, $F$ is the flux and, in the absence of source redshifts, we adopt the luminosity distance $d_L$ of the tentative host cluster.
Note that the weighted mean can mislead, as it is biased by higher-confidence measurements, which are skewed towards smaller values, and its small uncertainty does not reflect the large underlying dispersion;
a more careful study of the source properties is deferred to \S\ref{sec:Props}.

Panels b and c of Fig.~\ref{fig:Excess1} show a low-significance excess of \gama-ray sources extending throughout the $2.2<\tau<3.5$ range.
Consider the surface number density $\mathfrak{n}(\tau)$ of sources in decreasing radii $\tau$, starting far outside the clusters.
A gradual inward increase in $\mathfrak{n}$ towards the virial shock
is consistent with the expected elevated density of accreted \gama-ray sources near the cluster, in particular star-forming or starburst galaxies and AGN.
The elevated $\mathfrak{n}$ suggests that even far from the cluster, beyond $\tau\simeq 4$, nearby sources at redshifts close to that of the cluster make a non-negligible contribution to $\mathfrak{f}$, which thus cannot be strongly dominated by field, in particular background, objects.

The weak, $1.7\sigma$ localized peak at the $\tau\simeq 2.4$ virial shock (especially in panel c) may well be a statistical fluctuation on top of this gradual inward increase in $\mathfrak{n}(\tau)$.
On the other hand, this peak may include a component that is highly localized at the virial shock, as found in radio and X-ray catalog sources, tentatively attributed to galactic halos and outflows locally energized by the shock \citep{IlaniEtAl24a, IlaniEtAl24}.
The associated $L\gtrsim 10^{42}\erg\se^{-1}$ luminosities (see panel d) are comparable to those of the X-ray sources, and 1--2 orders of magnitude above the radio-source luminosities.

Unlike the low-significance excess of 4FGL-DR4 sources at or outside the virial-shock radius, the diminished number of sources just inside the virial shock is quite significant.
In particular, the absence of any significant 4FGL-DR4 sources in the $1.5<\tau<2$ range of the high-quality stacking analysis (panel c) is significant at the $3.4\sigma$ confidence level.
This result is consistent with the reported quenching of star-forming or starburst galaxies and AGN as they approach a cluster and cross into the ICM \citep[\eg][]{MahajanRaychaudhury10_ComaQuenching, PengEtAl10, vonderLindenEtAl10, WetzelEtAl12, HainesEtAl12, LiuEtAl19}.
However, unlike the gradual quenching reported previously, Fig.~\ref{fig:Excess1} shows that the effect is well localized just inside the virial shock.
Note a similar sharp drop, just inside the virial shock radius, in the normalized radial distribution of radio sources \citep[\eg LoTSS sources matched to SDSS galaxies by\,][]{deVosEtAl24}.
The strongly quenched $\mathfrak{n}$ supports the above conclusion that background objects cannot dominate $\mathfrak{f}$.

At smaller, $\tau<1.5$ radii, there is an excess of both diffuse emission and discrete sources.
Panel (a) of Fig.~\ref{fig:Excess1} shows a local peak in diffuse emission around $\tau\simeq 1.2$, discussed in \S\ref{sec:Diffuse}, and a strong central, $\tau<0.5$ diffuse excess, studied in {\PapII}.
Panels (b) and (c) show a strong excess of \gama-ray sources, which extends to the center of the cluster with local peaks around $\tau\simeq 1.1$ and $\tau\simeq 0.6$, considered next.

\section{Intracluster excess of 4FGL-DR4 sources}
\label{sec:ICM}

Figures \ref{fig:VSummary} and \ref{fig:Excess1} show a significant excess of 4FGL-D4 sources at $\tau<1.5$ radii, with comparable confidence levels found in both stacking analyses, as shown in Figs.~\ref{panel:b} and \ref{panel:c}.
In particular, for the larger, 205-cluster sample (panel b), we find a $4.3\sigma$ excess of $\tau<1.5$ (henceforth ICM) sources.
These 34 ICM sources are listed in \ref{app:Sample}, along with their tentative host clusters.
Excluding the central $\tau<0.5$ region still leaves a $3.5\sigma$ excess in the $0.5<\tau<1.5$ region, dominated by a $3.3\sigma$ narrow, $0.5<\tau<0.75$ peak and a $2.1\sigma$ broader, $1<\tau<1.5$ excess.
These nominal confidence levels are too conservative, however, as they are evaluated based on the non-quenched field.
Taking into account quenching inside the virial shock, for example by measuring the small $\FS>0$ in the $1.5<\tau<2$ region, yields a $>5\sigma$ excess of $0.5<\tau<1.5$ sources.

The two sources found in the central, $\tau<0.2$ region are established cluster-member AGN (both being high-significance, high-latitude, well-localized \gama-ray sources, classified as a quasar and a BL-Lac object in the SIMBAD database\footnote{\href{http://simbad.u-strasbg.fr/simbad/}{http://simbad.u-strasbg.fr/simbad/}}).
In addition, eight (four) ICM sources have optically-supported SIMBAD AGN classification in the panel b (panel c) analysis, constituting about $0.5\mathfrak{f}$ in the $\tau<1.5$ region.
Of these, five (two) sources were identified as high-redshift background AGN, alone accounting for $30\%$ ($20\%$) of $\mathfrak{f}$.
These numbers are consistent with the above arguments for a small fraction of background objects in $\mathfrak{f}$.
Notably, the $68\%$ ($95\%$) localization of these \gama-ray sources is $\gtrsim 0\dgrdot03$ ($\gtrsim 0\dgrdot05$), so most AGN associations are uncertain.
The source surface density excluding these putative AGN associations is shown (blue triangles) in panels (b) and (c).
The diminished ICM source excess remains significant even with respect to the initial $\mathfrak{f}$, and would become much stronger after correcting this field estimate by excluding all identified AGN in the ROI.

Many of the other ICM sources have, within their localization uncertainty, coincident radio counterparts, and were classified as blazar candidates \citep[\eg][]{FanEtAl16, ChangEtAl17, LefaucheurPita17, Chang19, AjelloEtAl22}.
If these \gama-ray- or radio-selected sources were indeed blazars, then the implied surface number density of AGN would rapidly increase inward; even the AGN fraction (defined with respect to member galaxies) would increase inwards \citep[\eg][]{HashiguchiEtAl23, deVosEtAl24}.
However, no such increase towards the centers of clusters was found in the surface number density of X-ray AGN \citep[except in the central $\tau<0.4$ of relaxed clusters;][]{RudermanEtAl05, YangEtAl09, EhlertEtAl2013, koulouridis2019high, NoordehEtAl20}, also implying that the AGN fraction decreases inwards.
Optically- or infrared-selected AGN have even smaller \citep[][and references therein]{MartiniEtAl07} and faster-declining \citep[\eg][]{GavazziEtAl11, PimbbletEtAl13, HashiguchiEtAl23} fractions.

Interestingly, an excess of X-ray sources was found to monotonically extend out to the virial radius in hard X-rays \citep[out to $\sim3R_{500}$ in massive, $M_{500}>10^{14}M_\odot$ clusters;][]{EhlertEtAl15}, and a $\tau\lesssim 1$ excess of X-ray sources was argued to be strong enough to raise the intracluster AGN fraction \citep[within $0.5<\tau<2$ in low, $M<10^{14}M_{\odot}$ mass clusters;][]{koulouridis2018xxl}; however, these sources may not be AGN: see \S\ref{sec:Discussion}.

This discussion, along with additional evidence presented below, suggests that most of the excess \gama-ray ICM sources may not, in fact, be blazars.
For instance, while the radial distribution of these sources does not agree with optical, IR, or X-ray AGN, it does roughly follow the distribution of a sample of shock relics, shown in the figure \citep[dot-dashed purple histogram in panels b and c, from table 2 of][]{NuzaEtAl17}.
Many of the 39 clusters hosting the 59 relics in the sample are not in MCXC, so we adopt $R_{500}=1\Mpc$ in the histogram, for simplicity and because this value coincides with the mean and median values for MCXC clusters with the high, $>10^{44}\erg\se^{-1}$ X-ray luminosities of almost all of these host clusters.
The histogram then shows the surface number density $\mathfrak{n}_r\equiv d N_{relics}/d\Omega'$ of shock relics stacked over these clusters.
The normalization of $\mathfrak{n}_r$, fixed by the sample size, is not meaningful here.

The excess ICM sources are clearly separated from the virial-shock and external regions, as little or no sources are found in the quenched, $1.5<\tau<2$ range.
Moreover, Fig.~\ref{panel:d} shows that the average properties of the ICM sources differ significantly from those in the field, as the former tend to be spectrally harder and less luminous.
ICM sources persist down to $\tau\lesssim0.5$, where hadronic emission was shown to be detectable even in a single massive, $z=0.023$ cluster (Coma), so it is interesting to ask if enhanced \gama-ray emission in dense regions, such as clumps and post-shock regions, may already appear in our low-redshift cluster sample as discrete 4FGL-DR4 sources, some incorrectly associated with AGN due to the poor localization.

None of the ICM sources belongs to the small group of 82 (out of 7195) sources classified in the catalog as extended, but even the brightest of this group has $>0\dgrdot12$ extent \citep[excluding two bright and compact pulsar wind nebulae;][and references therein]{FGL4DR4}, which in our sample corresponds on average to $\gtrsim 300\kpc$; the detectable extents of the much fainter ICM sources would be far larger.
The extended PSF thus yields large detectable extents and source localization uncertainties, typically exceeding the size of the cluster core, so the discrete nature of the sources does not rule out a diffuse hadronic origin.
Such sources would not in general have an optical counterpart, but sufficient magnetization would give rise to coincident radio emission, which could present as some type of radio relic; however, at low sensitivity or resolution, such a radio counterpart may appear compact, point-like, or not at all.

Figure \ref{panel:d} shows no strong evolution in the mean-weighted properties of 4FGL-DR4 sources across the $0<\tau<1.5$ range.
The mean luminosity does appear to decline monotonically with increasing $\tau$, which is somewhat more natural for an ICM origin than for blazars, but the significance is low and the dispersion is large; see \S\ref{sec:Props}.
As the figure shows, ICM sources show on average a flat, $s\simeq 2$ spectrum, which is consistent with hadronic emission from the flat, $du/d\ln E\simeq \const.$ CRI spectrum of {\PapII}, but is noticeably harder than found in field \gama-ray sources.
However, an average $s\simeq 2$ does not necessarily (but see \S\ref{sec:Props}) distinguish such hadronic emission from low-redshift blazars, because the latter also show similar mean spectra, as indicated for example by the two central sources dominating the $\tau<1/4$ bin in the figure.

In contrast, the characteristic $L\sim 10^{42} \erg\se^{-1}$ luminosities inferred for ICM sources in the figure are an order of magnitude or more lower than typically found in blazars with an optical association at the relevant low redshifts \citep{GhiselliniEtAl17, AjelloEtAl20}.
Such luminosities agree better with a locally elevated hadronic emission, as argued in \S\ref{sec:Diffuse} below.
Presently, suffice to note that for a flat spectrum, $L(0.1\mbox{--}100\GeV)\sim 10^{42} \erg\se^{-1}$ corresponds to $\nu L_\nu\simeq 1.4\times 10^{41}\erg\se^{-1}$, which is typical of bright radio minihalos \citep{GiacintucciEtAl19Expanding, TimmermanEtAl21}, giant halos \citep{vanWeerenEtAl19}, and shock relics \citep{PauloEtAl16Presentation, NuzaEtAl17} at $\nu =1.4\GHz$ frequencies (equivalent to a specific power $P_{1.4\mbox{\scriptsize{ GHz}}}\simeq 10^{25}\mbox{ W Hz}^{-1}$), and with radio phoenixes, which are soft, at $100\MHz$ \citep{Keshet25Phoenix}.

Finally, note that excess 4FGL-DR4 sources have no meaningful effect on the analyses of the virial-shock signal in {\PapI} or the diffuse central emission in {\PapII}.
Those analyses were verified with the updated, 4FGL-DR4 source-masked data around 75 clusters at $|b|>30\dgr$ latitudes, chosen, in addition, far from any significant well-localized source.
Only three catalog source are found in the virial-shock bin of these 75 clusters; all three are $>5\sigma$ significant, so all HEALPix pixels within $95\%$ PSF containment of each source were masked.
Only four sources are found within $\tau<1$ of these clusters, only two of which persist in the subset of 31 non-compact, $\theta_{500}>0\dgrdot25$ clusters used to model the central signal in {\PapII}.
Three of the four sources are significant and poorly ($a_{95}>0\dgrdot1$) localized, so the surrounding HEALPix pixels are again masked.
The fourth source is of low ($<5\sigma$) significance and marginal ($a_{95}\simeq 0\dgrdot08$) localization, so was not masked, but we verified that its presence does not noticeably modify the {\PapII} results.

\section{Properties of ICM sources}
\label{sec:Props}

Before analyzing in \S\ref{sec:InferredProps} the inferred properties of the excess ICM sources, it is useful for context to consider in \S\ref{sec:ExpectedProps} the various discrete hadronic \gama-ray sources expected in the ICM.

\subsection{Properties of expected hadronic sources}
\label{sec:ExpectedProps}

Excess hadronic \gama-ray emission is expected when the product of CRI and ambient nuclei densities is elevated.
The resulting enhanced pion production
leads to \gama-ray emission mainly from $\pi^0\to\gamma\gamma$ decay, but also from Compton scattering of cosmic microwave background (CMB) photons off the secondary CRE produced by $\pi^\pm$ decay.
The spectrum of the extended, central ICM CRI distribution is flat, $p\equiv 1-d\ln u/d\ln E\simeq 2$, according to radio observations and the \gama-ray analysis in {\PapII}.
The resulting $\pi^0$ decay signal has an $s\simeq 2$ spectrum, as does the spatially-integrated, combined Compton and synchrotron emission from secondary CRE.
Here and below, we refer to $>\GeV$ photons, beyond the turnaround energy of the $\pi^0$ signal; at lower energies, the LAT PSF becomes prohibitively extended for clusters, so is avoided in the diffuse analysis and is less relevant for discrete-source spectra.

CRE Compton losses dominate over synchrotron in weakly magnetized, $b\equiv B/(8\pi u_{cmb})^{1/2}\simeq (1+z)^{-2}B/3.2\muG<1$ regions, in which one expects a nearly pure $\mys\simeq 2$ spatially-integrated \gama-ray spectral index to result from the central CRI distribution.
Here, $B$ is the field amplitude, $u_{cmb}$ is the CMB energy density, and let us define a normalized $\epsilon_1\equiv \epsilon/\mbox{GeV}$ photon energy.
Compton \gama-ray emission is produced by CRE of high, $E\simeq 0.6\epsilon_1^{1/2}\TeV$ energy, which cool rapidly, over $\sim 2(1+b)^{-2}(1+z)^{-4}\epsilon_1^{1/2}\Myr$.
Hence, unlike radio synchrotron \citep{Keshet24}, the \gama-ray spectrum radiated by secondary CRE is weakly affected by CR diffusion and ICM dynamics.
The hadronic spectral index in such weakly-magnetized regions is therefore $\mys\simeq 2$ locally, and not only when integrated.

In more magnetized, $b\gtrsim1$ regions, the Compton spectrum is somewhat modified by CRE evolution.
In general, synchrotron losses soften the CRE and hence their Compton emission.
Some opposite, spectral hardening can manifest in bright regions from which older CRE diffuse away, but this effect is again small for rapidly cooling, high-energy CRE.
In either case, as the specific energy density injected by a flat CRI spectrum in $\pi^0\to\gamma\gamma$ photons is $\sim$twice higher than in their CRE counterparts \citep{KamaeEtAl06}, and most of the CRE energy is furthermore lost to synchrotron in these magnetized regions, modifications to the $\mys\simeq 2$ hadronic \gama-ray spectrum are small.

Consider the weak ICM shocks observed propagating outwards in some clusters, typically attributed to a merger event and sometimes giving rise to a shock relic.
A weak ICM shock compresses the gas by a factor $r_g=4\Mach^2/(3+\Mach^2)$, and CRI of a flat, $p=2$ spectrum by a larger factor $r_{cri}=\Mach^2$, so the product $r_g r_{cri}$ explains the enhanced brightness of the relics \citep{Keshet10}.
Here, $\Mach$ is the shock Mach number, and a $\Gamma=5/3$ adiabatic index is assumed, henceforth.
The compression of upstream secondary CRE, of $p=3$ steady state, can be even stronger, by a factor $4\Mach^2/(5-\Mach^2)$ formally diverging as $\Mach^2\to5$, beyond which diffusive shock acceleration (DSA) or shock modification are implied; however, such compression is quenched by radiative cooling even for radio-emitting CRE when CR diffusion is strong \citep{Keshet24}.

Indeed, the diffusion of CR in the ICM was found to be strong, of coefficient $D\gtrsim 10^{31}\cm^2\se^{-1}$.
If the compressed CRI are sufficiently confined in a $\lesssim 50\kpc$ shell trailing the shock before they diffuse away,
then the concentrated CRE they produce account for the spectrospatial properties of shock relics, including strong dimming and softening downstream \citep{Keshet24}.
In such a case, an edge-on shock would appear at sufficiently low resolution as a discrete \gama-ray source just behind the shock, due to the $\gtrsim r_g r_{cri}\simeq 4\Mach^4/(3+\Mach^2)$ enhancement in pion production within the shell.
This factor rises rapidly with Mach number, from $\sim9$ for $\Mach=2$ to $27$ for $\Mach=3$.

The associated luminosity increases with cluster mass, in proportion to the CRI energy density $u$ and the projected radiating volume.
A mass dependence $u(M_{500})$ was not yet demonstrated observationally, but adopting the $u\propto R_{500}^{-3}\int \dot{M}T\,dt\propto T\propto M_{500}^{2/3}$ dependence of a simple virial-shock injection model \citep{KeshetEtAl04}, where $T$ is the temperature, yields $L\propto M_{500}^{2/3}\propto L_{500}^{4/9}$, where the last step took the X-ray luminosity within $r<R_{500}$ as $L_{500}\propto M_{500}^{3/2}$ \citep[\eg][]{PrattEtAl09}.
A radiating volume $\propto R_{500}^2\propto M_{500}^{2/3}$ of an edge-on shell then gives $L\propto M_{500}^{4/3}\propto L_{500}^{8/9}$.
Shock-relic luminosities show a stronger mass dependence \citep{YuanEtAl15}, but here $B$ should be folded in.
As the ambient gas density declines rapidly with radius, farther \gama-ray sources of this type should in general become less luminous, consistent with Fig.~\ref{panel:d}, despite the slowly increasing $\Mach$ of outgoing shocks.

Shock relics are usually modeled, despite the challenges outlined below, by invoking the DSA of CRE.
DSA is well established in strong non-relativistic shocks, thought to yield
\begin{equation}\label{eq:DSA}
  p=2(\Mach^2+1)/(\Mach^2-1)
\end{equation}
CR spectra \citep{AxfordEtAl77, Krymskii77, Bell78, BlandfordOstriker78} when CR scattering is not too anisotropic \citep{keshet2020diffusive}.
However, DSA was not shown observationally or theoretically to be efficient in weak shocks.
Furthermore, DSA is quenched when perpendicular (to the shock normal) magnetic fields are strong or when CR diffusion is as strong as measured in the ICM, including in relic downstreams \citep{Keshet24}.

Indeed, evidence shows that most shock relics are not consistent with DSA.
In particular, their spectra are too far from DSA estimates \citep[given the measured $\Mach$;][and references therein]{vanWeerenEtAl19}, too universal, and too uniform along the shock front.
Instead, shock relics appear to share the same non-DSA origin as halos, as indicated directly
by systems capturing different stages in the smooth separation of a shock from a halo: shock still at the halo edge (no relic), shortly after separation (young relic with a gradually dimming halo--relic bridge), and at late times (non-detected bridge). For additional evidence and details, see \citet{Keshet10, Keshet24}.

DSA may however operate in the ICM if upstream conditions are sufficiently modified.
While no shock relic shows the integrated soft, pure power-law spectrum matching weak-shock DSA, phoenix relics present soft high-frequency spectra and were claimed to be tied to weak ICM shocks, although this claim is yet to be confirmed \citep[\eg][]{vanWeerenEtAl19}.
Phoenixes are widely modelled as aged CRE that recently underwent acceleration \citep{Kardashev62, JaffePerola73, KomissarovGubanov94} or compression \citep{EnsslinGopalKrishna01}, leading to increasingly curved spectra at high frequencies.
However, when sufficient high-frequency measurements are available, they are consistent with a pure soft power law; in fact, the entire spectrum is better fit in all phoenixes by hadronic emission from CRI of a pure power-law spectrum, in which case it is not necessary to invoke a recent violent event to re-energize a preexisting CRE population \citep{Keshet25Phoenix}.
The Compton counterpart of a radio phoenix may present as an extended nonthermal X-ray source \citep[\eg][]{BagchiEtAl98}.
A hadronic phoenix should show pronounced $\sim$keV Compton and $\sim$GeV $\pi^0\to\gamma\gamma$ emission; at higher photon energies, the signal would be strong only for the harder phoenixes \citep{Keshet25Phoenix}.

Next, consider a dense clump in the ICM.
As CRI diffuse and homogeneously permeate an $M_g\equiv 10^{12} M_{12}M_{\odot}$ gas mass, they give rise to hadronic signatures proportional to $M_g$.
Adopting the approximately constant $du/d\ln E$ energy density of CRI inferred from radio \citep{Keshet10} or from \gama-rays (\PapII), and an inclusive $\sigma\equiv 100\sigma_{100}\mbox{ mb}$ cross section for $\pi^0$ production by $E\simeq 100\GeV$ CRI \citep[\eg][]{KamaeEtAl06}, the resulting \gama-ray differential luminosity roughly becomes
\begin{equation}\label{eq:HaloGamma}
  \frac{dL}{d\ln\epsilon}(\epsilon\simeq 10\GeV) \simeq \frac{M_g}{m_p}c \sigma \frac{du}{d\ln E} \simeq 10^{41.0\pm0.5}\sigma_{100} M_{12} \erg\se^{-1}
  \coma
\end{equation}
where $m_p$ is the proton mass and $c$ is the speed of light.
A stronger signal is expected if the clump harbors a higher CRI energy density than inferred in the ICM, as suggested for example by the radio phoenix in cluster A1914 \citep{Keshet25Phoenix}.

The differential luminosity \eqref{eq:HaloGamma} corresponds to $\mys=2$ and $L\simeq 10^{41.8\pm0.5}\sigma_{100}M_{12}\erg\se^{-1}$, and should increase with cluster mass in proportion to $u\propto M_{500}^{2/3}$.
Compton and synchrotron emission from secondary CRE, jointly with $\sim$half the differential luminosity \eqref{eq:HaloGamma} and with the same spatially-integrated (but not necessarily local) $\mys\simeq2$ spectrum, may be observed as non-thermal X-rays and, in magnetized regions, as radio emission.
These sources may be enhanced by local CR, in general with a softer spectrum.
Star formation and AGN are thought to be quenched deep in the cluster, but fossil CRI could be important.

As a weak shock penetrates the low-pressure parts of the clump, a fraction $f_s$ of the gas mass is shocked, raising pion production by a factor $f_s r_g r_{cri}$; high pressure regions may be affected by compression modes.
In addition, if DSA is not quenched in such a clump, it can produce a soft CR power-law spectrum, possibly amplified by a fossil population.
Again, secondary CRE then give rise to extended, non-thermal X-ray emission, whereas $\pi^0$ decay can present as a \gama-ray source.

\subsection{Properties of excess 4FGL-DR4 sources}
\label{sec:InferredProps}

\begin{figure*}[t!]
    \centering
    \includegraphics[width=0.492\textwidth]{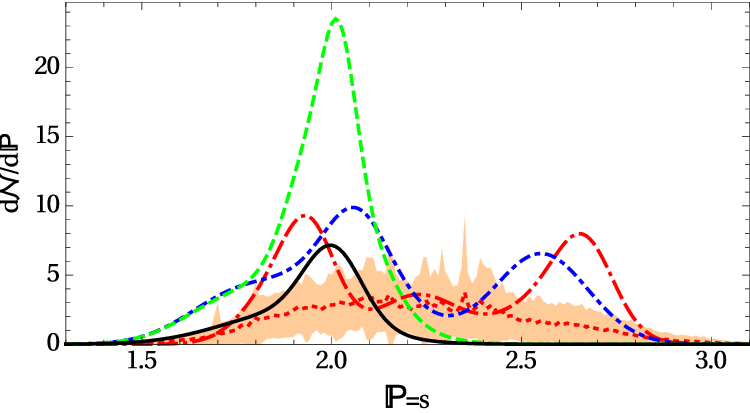}
    \includegraphics[width=0.492\textwidth]{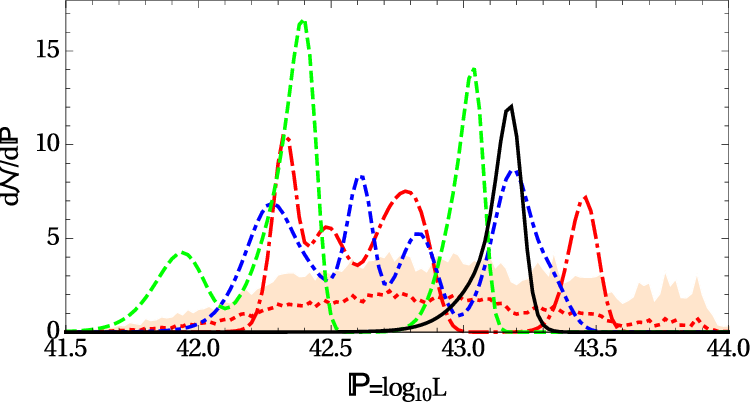}\\
    \includegraphics[width=0.492\textwidth]{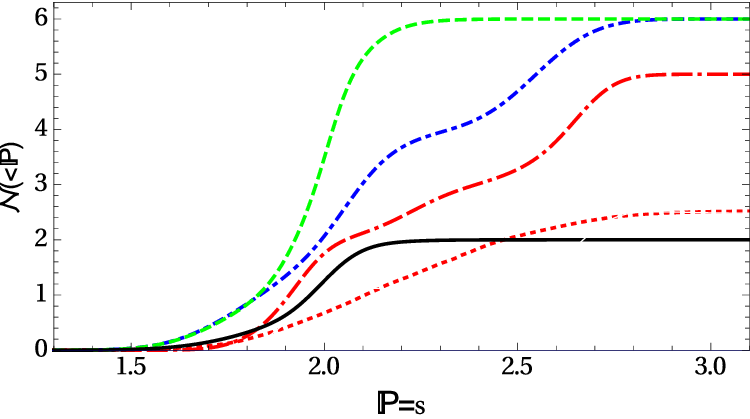}
    \includegraphics[width=0.492\textwidth]{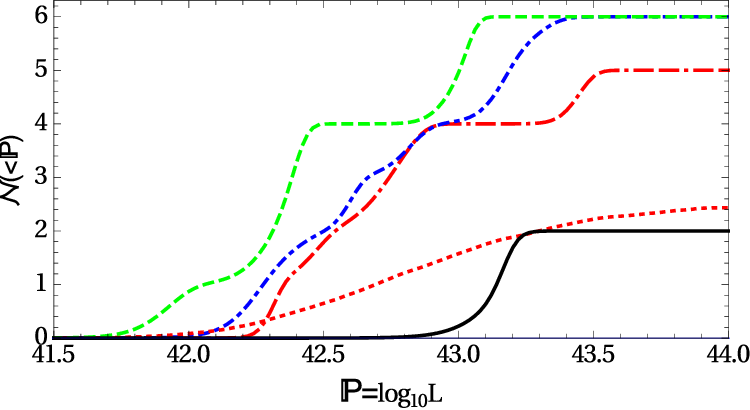}\\
    \includegraphics[width=0.492\textwidth]{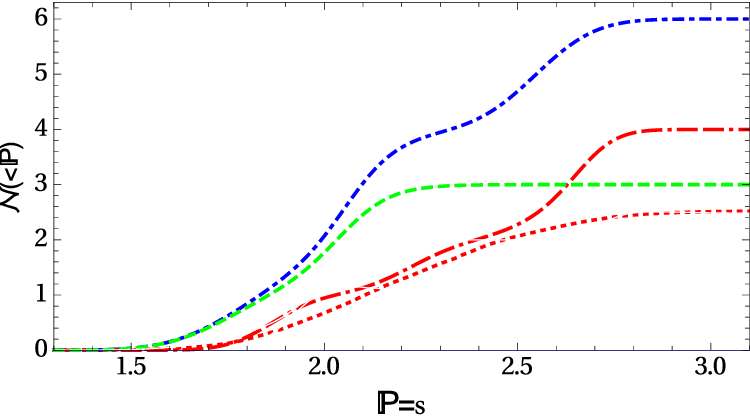}
    \includegraphics[width=0.492\textwidth]{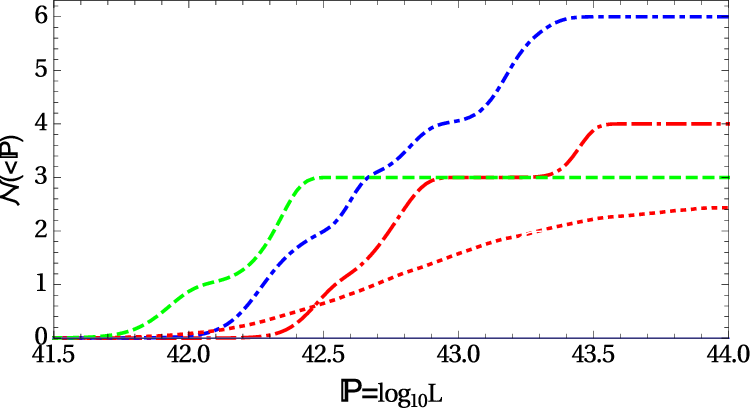}\\
    \caption{
        Spectral index $\mys$ (left column) and luminosity $L$ (right) differential (top row) and cumulative distributions among $\tau<1.5$ sources in the high-quality 4FGL-DR4 stacking (see Fig.~\ref{panel:c}), including (top and middle row) and excluding (bottom) sources with optical SIMBAD associations.
        Curves show excess sources in central ($\tau<0.2$ identified AGN; solid black curves), inner ICM peak ($0.5<\tau<0.75$; dot-dashed blue), outer ICM peak ($1<\tau<1.25$; double-dash-dot red), and non-peaked (combined $0.75<\tau<1$ and $1.25<\tau<1.5$; dashed green) regions, as well as sources in the field ($7<\tau<15$; dotted red with orange $1\sigma$ dispersion) with a conservative normalization (scaled to the solid angle of the outer ICM peak, not accounting for quenching).
    \label{fig:PropB30}
	}
\end{figure*}

In order to study different properties $\myP$ of the excess \gama-ray sources, Fig.~\ref{fig:PropB30} shows the differential $d\NS/d\myP$ and cumulative $\NS(<\myP)$ distributions of source spectra ($\myP=s$; left column) and luminosities ($\myP=L$; right column) in the high-quality stacking (corresponding to panel c in Fig.~\ref{fig:Excess1}); other samples, metrics, and details are deferred to \ref{app:Prop}.
Each panel demonstrates the properties of sources in different spatial regions, as well as the corresponding distribution among field ($7<\tau<15$; dotted red curves) sources.
Sources with optically-based SIMBAD identifications are removed in the bottom row, although not all optical associations are robust given the poor \gama-ray localization.

The figure confirms that the properties of $\tau<1.5$, \ie ICM sources, significantly differ from their field counterparts, as suggested by the weighted-mean profiles in Fig.~\ref{panel:d}.
In particular, a clear excess of spectrally flat, $s=2$ sources, found in all $\tau<1.5$ regions, is noticeably harder and narrower than in the field.
The significance of this excess depends on its spectrospatial definition, the quenching of the field, and any removal of identified sources.
For instance, the $1.75<\mys<2.25$ excess at $0.5<\tau<0.75$ is significant at the $4.9\sigma$ level if the coincident field is diminished by a factor of $2$, even after all optically based SIMBAD-identified sources are removed.
This estimate is conservative, as the combined quenching and removal of identified sources also in the field should eliminate most of $\mathfrak{f}$.

A secondary excess of soft sources peaks at $s\simeq 2.6$; under the above assumptions, including a conservative $0.5\mathfrak{f}$ field, it is significant at the $3.2\sigma$ level.
Consequently, the overall distribution of ICM source spectra, both in this $0.5<\tau<0.75$ peak and in the $\tau<1.5$ region overall, is bimodal, unlike the unimodal, broad field distribution.
The soft sources are found exclusively in the $\tau\simeq 0.6$ and $\tau\simeq 1.1$ peaks.
Hard sources outnumber soft sources in the ICM region by a factor of $\sim2$;
among the ten sources with optical associations, only two are soft.
Excluding such associations leaves 14 hard and 10 soft sources when using an arbitrary $\mys_{th}\equiv2.25$ threshold (chosen in the middle of the $2.1<\mys<2.4$ range, which contains only three sources).

The discussion in \S\ref{sec:ExpectedProps} suggests that the hard sources may be dense clumps permeated by cluster CRI, whereas the softer sources may be clumps with their own, softer CRI population; in both cases, a weak shock may be needed to boost the hadronic signal.
Alternatively, the hard sources may be the \gama-ray counterparts of shock relics, whereas the rarer, soft sources possibly arise from $\pi^0$ decay in phoenix relics.
Shock relics are in general more abundant than phoenix relics; for example, the latter constitute only $\sim 15\%$ of the \citet{NuzaEtAl17} sample.
However, at the low, $z<0.1$ redshifts of clusters in the present sample, the phoenix fraction doubles (when including candidates) or triples (otherwise).

Radio relics require strong magnetization, present a low surface brightness, and are uncommon; shock relics were mostly reported at redshifts higher than those of clusters in the present sample, whereas phoenixes with $p<4$ are rare \citep{Keshet25Phoenix}.
Hence, we cannot reliably test if the \gama-ray sources in our sample coincide with radio relics.
Only Abell S753 shows both an ICM 4FGL-DR4 source and a known radio relic, but the latter is a bright phoenix candidate near the center of the cluster, whereas the \gama-ray source is more peripheral.
One 4FGL-DR4 source is projected in the ICM of A2063, which is often mentioned as hosting a relic near its center, but the latter should probably be classified as a radio galaxy \citep{GeorgeEtAl17}, and in any case is far from the \gama-ray source.

\begin{figure}[t!]
    \centering
    \includegraphics[width=0.45\textwidth]{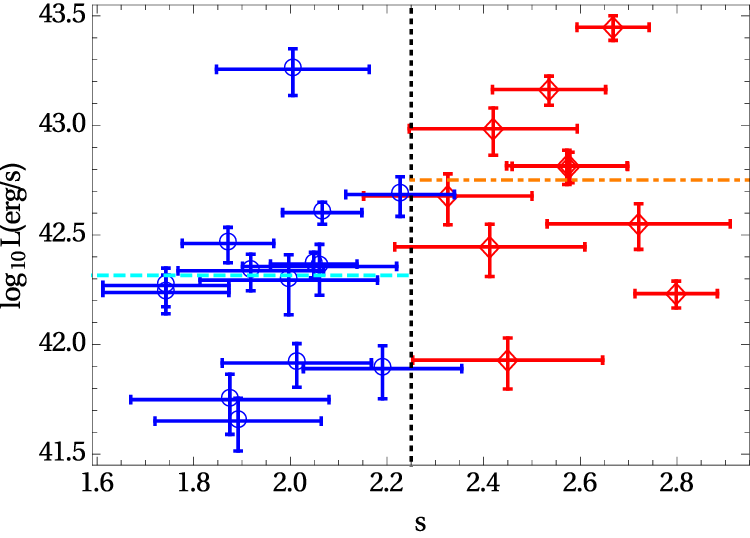}
    \caption{
        Luminosity vs. spectrum for ICM ($\tau<1.5$) 4FGL-DR4 sources without SIMBAD optical associations.
        Softer sources (red diamonds) are on average somewhat brighter than hard sources (blue circles), as demonstrated by the respective median luminosities (dot-dashed and dashed horizontal lines), but the dispersion is large and no clear scaling is identified.
    \label{fig:LvsS}
	}
\end{figure}

ICM sources are on average fainter than their field counterparts.
The right column of Fig.~\ref{fig:PropB30} shows, for instance, that $80\%$ of the excess source distribution lies in the $L<8\times 10^{42}\erg\se^{-1}$ luminosity range, compared to only $56\%$ in the field (at the tentative cluster redshift).
The soft sources are slightly more luminous, on average, than their harder counterparts, as illustrated in Fig.~\ref{fig:LvsS}, where optical associations are excluded.
For example, the median or weighted-mean luminosities of $\mys>\mys_{th}$ sources are about double those of $\mys<\mys_{th}$ sources.
While the dispersion in \gama-ray source luminosities is considerable,
it is much smaller than expected in blazars \citep{GhiselliniEtAl17}.
In addition, for the relevant $L$ range, the spectral distribution differs significantly from blazars and radio galaxies, which show much larger dispersion around $\mys=2$, and no bimodality or soft excess \citep{AjelloEtAl20}.

\begin{figure}[h]
    \centering
    \includegraphics[width=0.45\textwidth, trim={0 0.4cm 0 0}, clip]{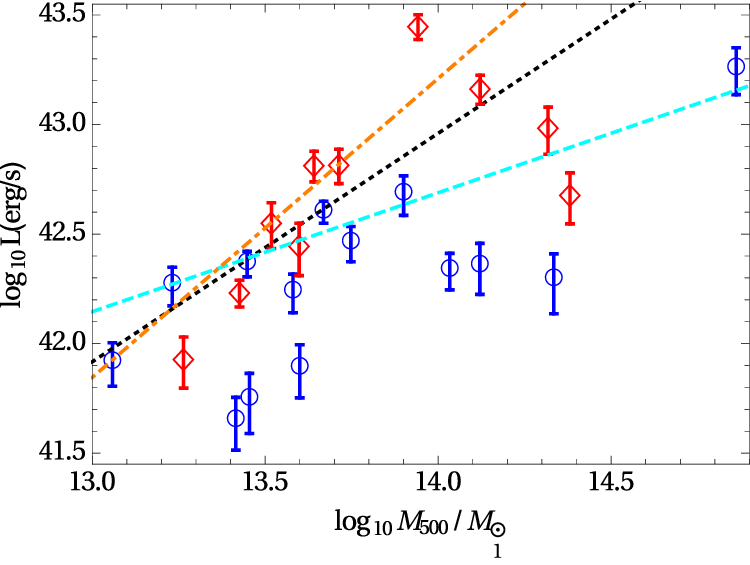}
    \caption{
        Source luminosity vs. host cluster mass,
        for the same sources shown in Fig.~\ref{fig:LvsS}.
        Positive $L$--$M_{500}$ correlations are found among hard (dashed cyan), soft (dot-dashed orange), and all (dotted black) sources.
    \label{fig:LvsM}
	}
\end{figure}

The \gama-ray luminosities $L$ of ICM sources correlate with the $M_{500}$ masses and $L_{X,500}$ X-ray luminosities of their host clusters.
Figure \ref{fig:LvsM} plots $L$ vs. $M_{500}$ after excluding optical associations.
The Spearman ranked correlation coefficient $\rho=0.68$ indicates a significant ($p\mbox{-value}=0.0002$) correlation.
The best fit (dotted black line) is linear, $L\propto M_{500}^{1.0\pm0.1}$.
For the hard (soft) sub-sample, $\rho=0.62$ ($\rho=0.76$), and the mass dependence is weaker (stronger), $L\propto M_{500}^{0.5\pm0.2}$ ($L\propto M_{500}^{1.4\pm0.2}$).
The correlations between source \gama-ray luminosities and cluster X-ray luminosities are equally strong, with expectedly smaller slopes; in particular, $L\propto L_{X,500}^{0.6\pm0.1}$ for hard and soft sources combined.

The strong correlations between ICM source luminosities and their host cluster masses is natural for emission associated with the ICM, in particular if these sources are the \gama-ray counterparts of radio relics.
In such a case, the $L\propto M_{500}^{4/3}$ relation, expected for CRI injected in the simple virial-shock model in \S\ref{sec:ExpectedProps}, is not too far from the correlation observed.
In contrast, a \gama-ray luminosity--host mass correlation is not expected for compact objects residing in clusters.
In particular, AGN luminosities show no significant dependence upon host mass, with a weak anti-correlation \citep[\eg][and references therein]{PowellEtAl20,AirdCoil21}, and radio-galaxy luminosities show no host-mass dependence \citep[except for the central galaxy, which we exclude;][]{SommerEtAl11}.

\section{Diffuse $r\sim R_{500}$ excess: hadronic model}
\label{sec:Diffuse}

Any local ICM disturbance that locally raises the gas density by some factor $r_g$ or amplifies the $du/d\ln E=10^{-13.6\pm0.5}$ erg cm$^{-3}$ CRI energy density by some factor $r_{cri}$, for example in clumps or behind weak shocks, would raise the \gama-ray emissivity by a factor $r_{inj}=r_g r_{cri}$, generally leading to a local excess in the diffuse signal.
Similarly, if the ICM \gama-ray sources discussed in \S\ref{sec:ICM} and \S\ref{sec:Props} are indeed excess hadronic emission, then in some cases such an excess may not be identified as sufficiently significant and localized for a discrete source, contributing, instead, to the diffuse signal.
In particular, discrete sources arising from edge-on weak shocks imply a diffuse enhancement by shocks in less favorable projections.

Indeed, even after masking all 4FGL-DR4 sources, a diffuse excess still remains in the ICM region, distinct from the strong diffuse signals in the central ({\PapII}) and virial-shock ({\PapI}) regions.
This $2.6\sigma$ excess peaks around $\tau\simeq 1.2$ in figures \ref{fig:VSummary} (magenta diamonds) and \ref{panel:a}, which are based on stacking the nominal sample of 75 clusters with $\theta>0\dgrdot2$ at $|b|>30\dgr$.
In those figures, only significant, $>5\sigma$ sources are masked, but masking also all non-significant 4FGL-DR4 sources has a negligible effect on the results.

\begin{figure}[b!]
    \centering
    \includegraphics[width=0.5\textwidth]{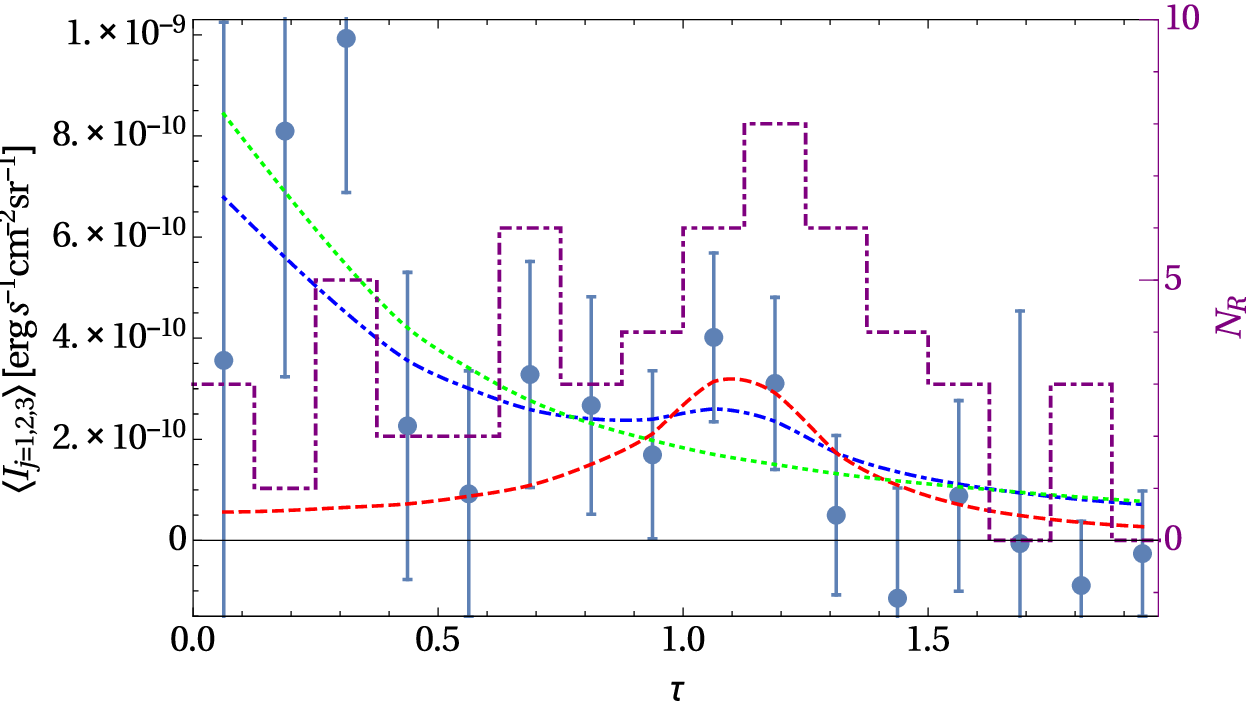}
    \caption{
        Stacked LAT brightness for the non-compact cluster sample, averaged over energy bands 1--3 (error bars with left axis).
        Curves show best fit models for the central excess ($\tau<0.5$; dotted green), ICM excess ($\tau\simeq 1.1$; dashed red), and the full cluster region ($\tau<2$; dot-dashed blue).
        A histogram of shock relics (dot dashed purple, with right axis) is shown as in Fig.~\ref{fig:Excess1}, but for relic counts $N_R$ rather then their surface density.
    \label{fig:Diffuse1}
	}
\end{figure}

In order to analyze this intermediate $\tau$ signal, referred to as the ICM excess, in conjunction with the central cluster excess, we use the smaller sample of 31 non-compact, $\theta_{500}>0\dgrdot25$ clusters, for which the central signal was modelled in {\PapII}.
Figure \ref{fig:Diffuse1} shows the brightness profile stacked over this sample, averaged over energy channels 1--3.
The high-energy channel 4 is excluded in the figure because it has insufficient counts in this small cluster sample with the required high, $\Delta\tau=1/8$ radial resolution.
As the figure shows, the excess approximately coincides with the maximum in the shock-relic count histogram.

Unlike the central or virial diffuse signals, or the discrete ICM source excess, the diffuse ICM excess is not highly significant on its own accord.
Its low, $\sim2.6\sigma$ significance, and the poor statistics given the small cluster sample and high resolution, yield prohibitively large uncertainties in the linear regression of individual channels.
Modeling the 1--3 channel-averaged signal suffices, however, for estimating the compression factor needed to explain the ICM excess as hadronic, and the implied Mach number of a corresponding shock.

Following {\PapII}, the specific brightness is modelled as
\begin{equation}\label{eq:HModel}
  \!\!\!I_\epsilon(\bmt) = \! \frac{c \epsilon'}{4\pi(1+z)^3} \!\! \int\!\! d\boldsymbol{\Omega} \, P(\bmp) \!\! \int\!\! dl \, n(\bmr) \!\! \int \!\! dE \frac{d^2u(\bmr,E)}{dE^2}\,  \frac{d\sigma_\gamma(E,\epsilon')}{d\epsilon'}\coma \!\!
\end{equation}
where
$\epsilon'=(1+z)\epsilon$ is the radiated photon energy,
$\bmp$ is the angular separation between observed ($\bmt$) and integrated ($\boldsymbol{\Omega}$) sky directions,
$P(\bmp)$ is the LAT PSF in the direction of the cluster,
and $l$ is the line-of-sight coordinate.
The differential inclusive cross-section
$d\sigma_\gamma(E,\epsilon')/d\epsilon'$ for $\epsilon'$ photon production is computed using the formulation of \citet[][with corrected parameters and cutoffs; T. Kamae \& H. Lee, private communications 2010]{KamaeEtAl06}.

Following the results of {\PapII}, the stationary ICM is modelled with a homogeneous and spectrally-flat CRI distribution, and with the default MCXC gas distribution following the \citet{PrattArnaud02} AB model with number density
\begin{equation}\label{eq:nMCXC}
  n\propto \left[1+\left({\tau}/{\tau_c}\right)^2\right]^{-\frac{3\beta}{2}+\frac{\alpha}{2}} \tau^{-\alpha} \coma
\end{equation}
with parameters $\alpha=0.525$, $\beta=0.768$ \citep{PiffarettiEtAl11}, and a scaled $\tau_c=0.1$ core radius corresponding to the models available for the individual clusters in the sample.
The ICM excess is modeled as an $r_{inj}$ enhancement in the rate of pion injection into the gas, localized at $\tau_0-\delta\tau<\tau<\tau_0+\delta\tau$ radii, motivated \eg by an edge-on shock as found in shock relics.

Due to the poor statistics, the best multi-channel fit gives a highly uncertain $p=1.8\pm5.5$ CRI spectral index for the enhanced emission.
Here and below, uncertainties indicate $68\%$ containment in a conservative, multivariate fit.
Fixing a flat, index $p=2$ spectrum for the ICM excess then gives best-fit values $\tau_0\simeq 1.1$, $\delta\tau\simeq 0.1$, and a small compression factor corresponding to an $\Mach<2$ shock, but the uncertainties are still prohibitively large.
Fixing $p=2$, $\tau=1.1$, and $\delta\tau=0.1$ yields $r_{inj}=3.9\pm21.6$, corresponding to a better-determined, $\Mach=1.5_{-0.5}^{+1.4}$ shock.

Better results are obtained when adopting a flat, $p=2$ CRI spectrum in both central and ICM regions, and fitting the brightness averaged over the 1--3 energy channels instead of in each channel separately.
The best fit gives $\tau_0\simeq 1.1$, $\delta\tau\simeq 0.1$ and $\Mach\simeq 1.8$, but the multivariate uncertainties are still prohibitively large.
Fixing $\tau=1.1$ and $\delta\tau=0.1$ gives $r_{inj}=7.0\pm14.0$, corresponding to $\Mach\simeq 1.8_{-0.8}^{+0.9}$ (dot-dashed blue curve in Fig.~\ref{fig:Diffuse1}).
Separately fitting the central (dotted green) and ICM (dashed red) regions yields a better constrained $r_{inj}=12.5\pm3.8$, corresponding to $\Mach\simeq 2.2_{-0.2}^{+0.3}$.
Similar results with somewhat lower $\Mach$ are obtained if we retain the $j=4$ channel in the averaging, despite its poor statistics.

Although these linear regression variants carry large uncertainties, they suffice to demonstrate that the compression induced by a weak, $\Mach\sim2$ shock can locally enhance pion production to the level needed to account for the diffuse ICM excess.
The results are not limited to a shock, as dense clumps or bubbles of enhanced CRI density can give rise to the same $r_{inj}$.
For instance, in an adiabatically compressed clump, the CRI energy density rises by a factor $r_g^{(p+2)/3}$, so the above $r_{inj}\simeq10$ corresponds to $r_g\simeq 2.7$ adiabatic gas compression.
For a bubble within an ICM region with an initial $f_0=1\%$ ratio of CRI-to-gas pressures, reaching $r_{inj}=10$ at isobaric ($1+f_0=r_g+r_{cri}f_0$) and isothermal balance requires $r_g\simeq 0.9$; here, full gas evacuation leading to no pion production sets at $r_{inj}\geq(1+f_0)^2/(4f_0)\simeq 26$.

Regardless of modelling, the $100\MeV$--$100\GeV$ luminosity corresponding to the diffuse ICM excess is on average $L\simeq 3\times 10^{41}\erg\se^{-1}$ per cluster, for median sample parameters.
This estimate is broadly consistent with the excess 4FGL-DR4 sources, if the numbers of cataloged vs. non-cataloged sources are comparable: while the excess ICM sources in the catalog are on average $\sim10$ times brighter than this $L$ estimate, there is only one such source per $\sim 8$ clusters, and only one source in the $1.0<\tau<1.25$ region per $\sim 30$ clusters.

\section{Summary and discussion}
\label{sec:Discussion}

This third paper in a series, which is devoted to stacking \gama-ray data around MCXC clusters scaled to their $R_{500}$ radii, stacks and studies both discrete 4FGL-DR4 sources, and diffuse \emph{Fermi}-LAT \gama-ray emission.
The results shed light not only on cataloged sources inside and around the clusters, but also on the $2.6\sigma$ diffuse $\tau\simeq 1.1$ excess previously found between the virial, $\tau\simeq 2.4$ excess of {\PapI} and the central, $\tau\lesssim0.5$ excess of {\PapII}.

A modest, gradual inward increase in source density towards the virial shock, with a low-significance, $1.7\sigma$ peak at the shock itself, is followed by a strong, $3.4\sigma$ quenching of these sources in the $1.5<\tau<2$ region, as seen in Figs.~\ref{fig:VSummary} (gray six-pointed stars) and \ref{fig:Excess1} (panels b and c).
Such a behavior is broadly consistent with previous studies of accreted star-forming or starburst galaxies and AGN.
However, the results indicates that quenching is more rapid than previously thought, taking place just inside the virial shock, within a few $100\kpc$.

Such rapid quenching is at tension with the slow ram-pressure stripping \citep[][and references therein]{Boselli22} or cold-gas strangulation \citep{LarsonEtAl80, PengEtAl15} processes, usually invoked to explain the suppressed star formation and AGN activity in the ICM over $\gtrsim \Gyr$ timescales; note, however, that accretion through cold streams \citep{KeresEtAl05, DekelBirnboim06} should be quenched rapidly.
Some of the excess $\tau \simeq 2.4$ sources may be the \gama-ray counterparts of the coincident excess radio and X-ray catalog sources \citep{IlaniEtAl24a, IlaniEtAl24}, tentatively attributed to galactic halos and outflows energized by the shock, thus explaining away part of the fast quenching.

Deeper inside the clusters, a strong excess of discrete sources is found throughout the $\tau<1.5$ region (Figs. \ref{fig:VSummary}, \ref{panel:b}, and \ref{panel:c}).
This excess is nominally at the $4.3\sigma$ confidence level, based on the non-quenched field, but is $>5\sigma$ when accounting for even conservative quenching.
Of the 34 excess sources (listed in \ref{app:Sample}), found in the ICM regions of 205 clusters, two are member AGN in the centers of clusters, and eight others have optical associations classifying them as (background, mostly) blazars in the SIMBAD database; however, such associations are not all robust, given the poor \gama-ray localization.
Most of the remaining sources were tentatively categorized as blazar candidates, but this classification is challenged by the results.

The excess $\tau<1.5$, so-called ICM sources show properties significantly different from those of field ($\tau>7$) sources, as seen in Figs.~\ref{panel:d} and \ref{fig:PropB30}--\ref{fig:LvsM}.
Their radial distribution (Figs.~\ref{panel:b}, \ref{panel:c}), bimodal spectral distribution with a narrow $\mys\simeq 2$ component and a smaller $\mys\simeq 2.6$ peak (Fig.~\ref{fig:PropB30}, left), luminosity distribution strongly correlated with cluster mass (Fig.~\ref{fig:LvsM}) and concentrated in the narrow, $L\simeq(1\mbox{--}8)\times 10^{42}\erg\se^{-1}$ range (Fig.~\ref{fig:PropB30}, right), along with additional evidence outlined below, all agree with hadronic clumps or the \gama-ray counterparts of radio relics, better than with compact sources such as AGN or radio galaxies.

Indeed, such excess sources, as well as the diffuse $\tau\simeq 1.1$ excess seen in Figs.~\ref{panel:a} and \ref{fig:Diffuse1}, can be accounted for by weak, $\Mach\simeq 2$ shocks or comparable ICM substructure or disturbances (see \S\ref{sec:ExpectedProps} and \S\ref{sec:Diffuse}).
The main, hard, $\mys\simeq 2$ component of the source distribution is consistent with excess pion production in hadronic clumps or shock relics, themselves natural consequences of shock compression, given the spectrospatially flat CRI distribution, which was predicted based on radio observations and established by \gama-rays in {\PapII}.
The secondary, soft, $\mys\simeq 2.6$ component may consist of the \gama-ray counterparts of phoenix relics or some similar form of excess pion-production, for example in hot bubbles, which is not dominated by the pervading $p\simeq2$ population of cluster CRI.

The $0.2\lesssim\tau<1.5$ excess 4FGL-DR4 sources are better explained by ICM substructure or radio-relic counterparts than by AGN or radio galaxies, for several reasons:
\begin{enumerate}[leftmargin=*, itemsep=-3pt]
\item
After the strong quenching of \gama-ray galaxies and AGN found inside the virial shock, presumably by ICM interactions, a subsequent sharp rise in their density would not be natural.
\item
Indeed, optical and X-ray selected AGN show no such trend outside the very central, $\tau<0.4$ region (see \S\ref{sec:ICM}).
\item
In contrast, the radial distributions of the sources (Fig.~\ref{fig:Excess1}) and diffuse emission (Fig.~\ref{fig:Diffuse1}) broadly agree with shock relics.
\item
No associated optical or radio members, as expected in AGN or radio galaxies, are identified coincident with the sources, although the well-observed host clusters lie at low redshifts.
\item
The strong correlation between source luminosity and cluster mass is natural for hadronic emission, and typical of shock relics, but is neither expected nor found in AGN or (non-central) radio galaxies (Fig.~\ref{fig:LvsM} and related discussion).
\item
The $L\simeq (1\mbox{--}8)\times 10^{42}\erg\se^{-1}$ luminosity range of most sources is comparable to bright radio relics, but too low and tightly distributed for blazars (Fig.~\ref{fig:PropB30} and related discussion).
\item
The source luminosity declines, on average, with increasing $\tau$ (Fig.~\ref{fig:Excess1}), which is natural for the \gama-ray counterparts of shock relics, but not generally expected in AGN or galaxies.
\item
The bimodal $\mys$ distribution (Fig.~\ref{fig:PropB30}) is not typical of AGN, radio galaxies, or their combination, whereas radio relics do come in two variants of markedly different spectra (\S\ref{sec:ExpectedProps}).
\item
The main, $\mys\simeq 2$ spectral component is too narrow (Fig.~\ref{fig:PropB30}), given the low $L$, for both blazars and radio galaxies.
\item
The diffuse excess (Figs.~\ref{fig:Excess1}a and \ref{fig:Diffuse1}) is more natural for the counterparts of relics, characterized by extended emission of low surface brightness, than for blazars or galaxies.
\end{enumerate}

Some of the clusters harboring ICM sources are merger clusters, while others show cool cores; no extended features are recognized as strongly associated with the \gama-ray sources.
For example, although Coma is excluded from the sample due to its large sky extent, the $\tau\simeq 1.0$ source 4FGL J1256.9+2736 southwest of its center is of interest, as it resembles the soft excess ICM sources, with $s=2.68\pm0.14$ and $L=(2.1\pm0.4)\times 10^{42}\erg\se^{-1}$.
The source is poorly ($68\%$ ellipse: $0\dgrdot13$) localized, overlapping the radio bridge to the southwest, but also the $\Mach\simeq 2$ shock extending south-through-west of the cluster \citep{PlanckComa12, UchidaEtAl16}.

Hadronic emission from weak shocks and ICM substructure should have non-thermal X-ray and, where magnetization is sufficiently strong, also radio, counterparts; however, the low surface brightness challenges their detection.
Interestingly, although optical, IR, and X-ray AGN do not typically show a strong $0.4<\tau<1.5$ excess, an excess of X-ray sources was reported in this region \citep{koulouridis2018xxl}, mainly $10^{42}\erg\se^{-1}<L(0.5\mbox{--}2\keV)<10^{43}\erg\se^{-1}$ sources with a radial overdensity profile similar to that in Fig.~\ref{fig:Excess1}.
Although these \emph{XMM} sources were matched with spectroscopically selected galaxies, the robustness of this association is unclear.
In particular, the sources presented in figure 7 therein, around clusters XLSSC 041 and XLSSC 561, appear in \emph{XMM} images as extended, and possibly elongated perpendicular to the cluster radius.
Such sources may thus be relic counterparts, and not AGN related.
Similarly, at least some of the \emph{Chandra} sources of \citet[][listed in table 5 therein]{EhlertEtAl15} are extended in \emph{Chandra} images, well beyond their associated optical galaxy.

\paragraph*{Acknowledgements} \begin{small}
I am grateful to Y. Lyubarsky, K.~C. Hou, I. Gurwich, A. Ghosh, and the late G. Ilani, for helpful discussions.
This research was supported by the Israel Science Foundation (ISF grant No. 2126/22), and used the SIMBAD database, operated at CDS, Strasbourg, France.
\end{small}

\bibliographystyle{elsarticle-harv-author-truncate}
\bibliography{Virial}

\appendix

\section{Sensitivity tests}
\label{app:Sensitivity}

The diffuse LAT-stacked signals are not sensitive to the precise choice of analysis parameters and selection cuts, as discussed in detail in {\PapI} and corroborated in {\PapII}.
Here, we verify that the modest $\tau\simeq 1.1$ signal is not sensitive to the choice of clusters or analysis parameters, either. As this diffuse excess coincides with an excess of discrete sources, we verify that the diffuse signal is not significantly affected by the choice of point-source masking. In particular, the same excess is found when masking all 4FGL-DR4 sources and when masking only significant, $>5\sigma$ sources.

\begin{figure*}
    \centering
    \includegraphics[width=0.45\textwidth]{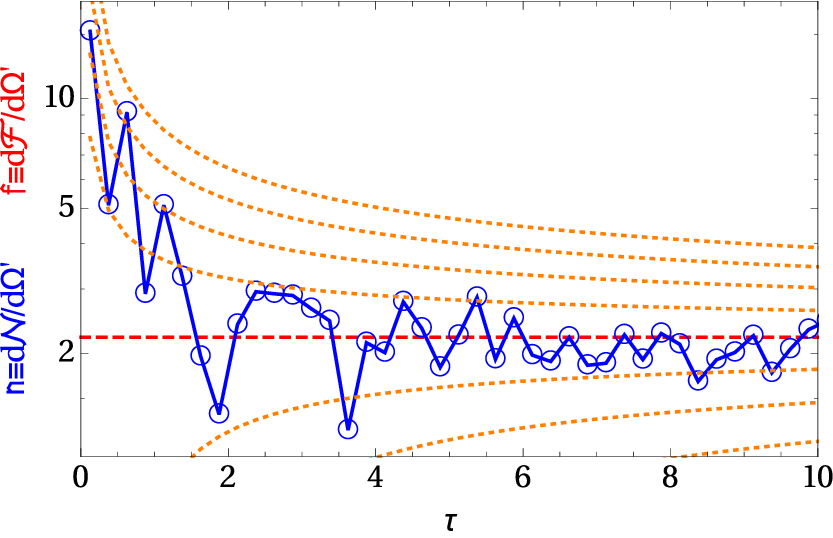}
    \includegraphics[width=0.45\textwidth]{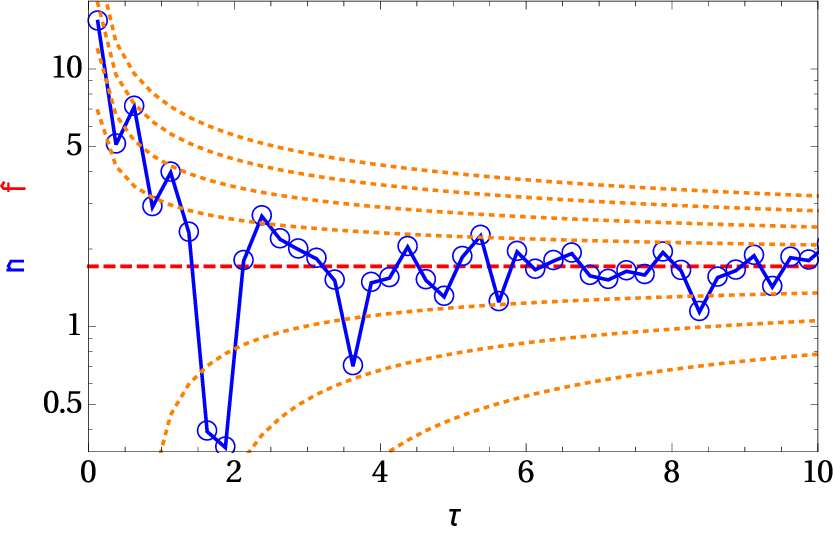}\\
    \includegraphics[width=0.45\textwidth]{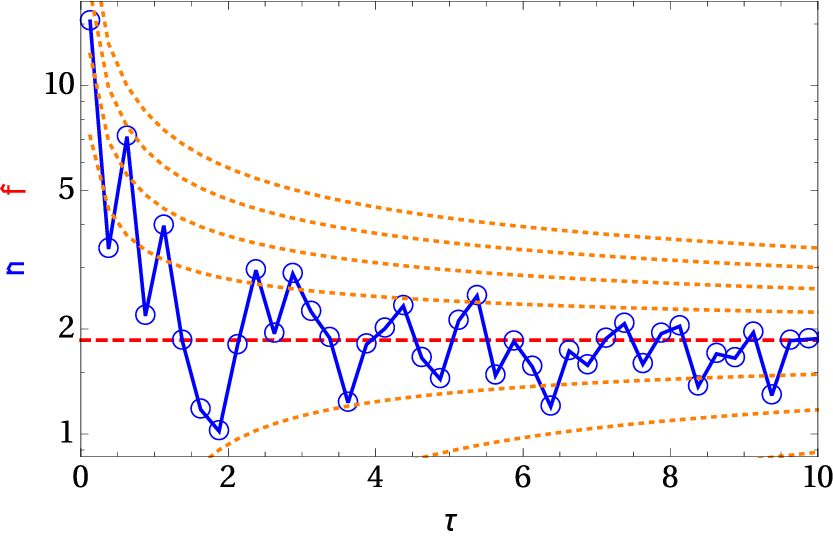}
    \includegraphics[width=0.45\textwidth]{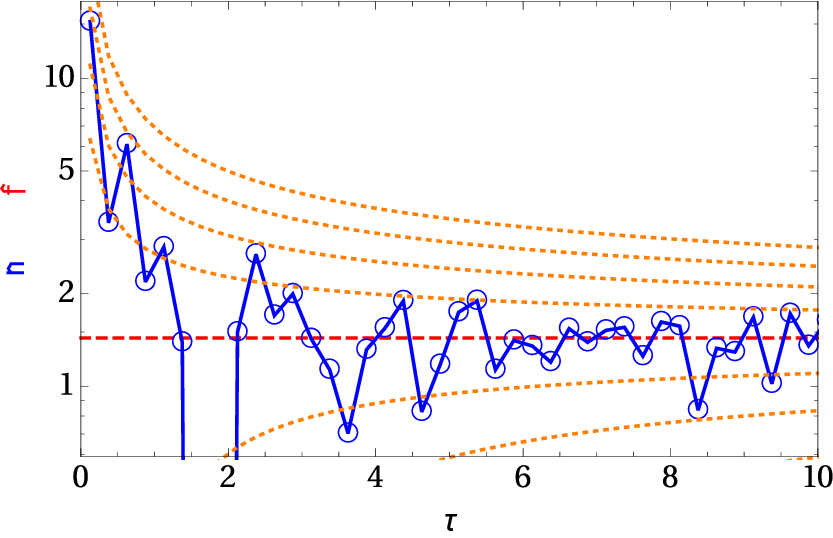}\\
    \includegraphics[width=0.45\textwidth]{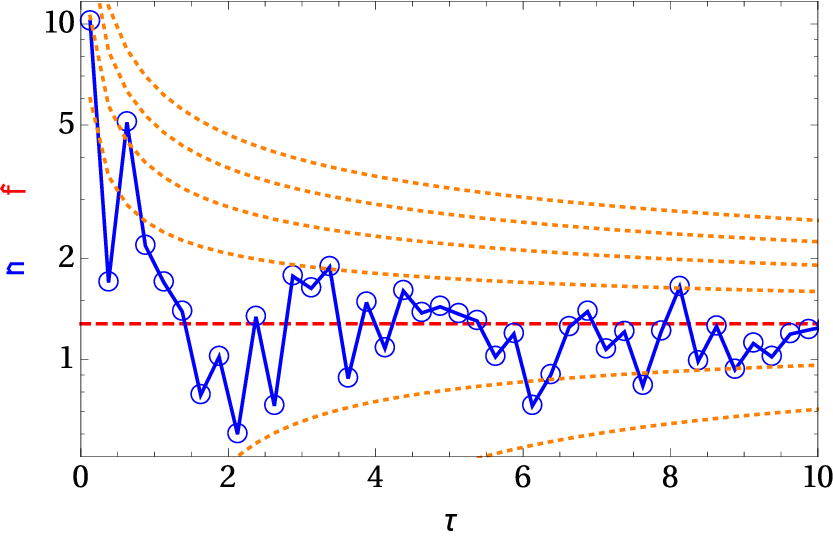}
    \includegraphics[width=0.45\textwidth]{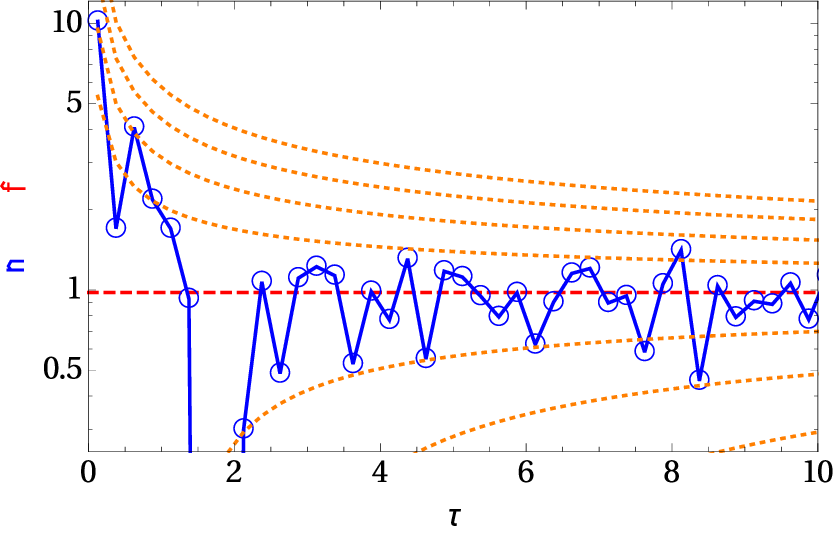}\\
    \caption{
        Same as Fig.~\ref{fig:Excess1} (panels b and c), but for $|b|>20\dgr$ (left column) or $|b|>30\dgr$ (right), for all (top row), only significant ($>5\sigma$; middle), or only significant and well localized ($<0\dgrdot08$; bottom) 4FGL-DR4 sources.
    \label{fig:Psc1}
	}
\end{figure*}

The presence of a strong excess of 4FGL-DR4 sources in the $\tau<1.5$ region is robust, recovered for various modifications of the analysis parameters and cluster selection cuts. However, as this excess consists of only a few dozen sources, its detailed profile and properties do depend on the analysis parameters. Figure \ref{fig:Psc1} demonstrates the 4FGL-DR4 excess for different cuts of latitude, source significance, and source localization. Sources at larger radii appear to be poorly localized; to wit, the $\tau\simeq 1.1$ discrete source excess vanishes when only well-localized, $a_{95}<0\dgrdot08$ sources are taken into account.

\section{Source properties}
\label{app:Prop}

\begin{figure*}[t!]
    \vspace{-2cm}
    \centering
    \includegraphics[width=0.492\textwidth]{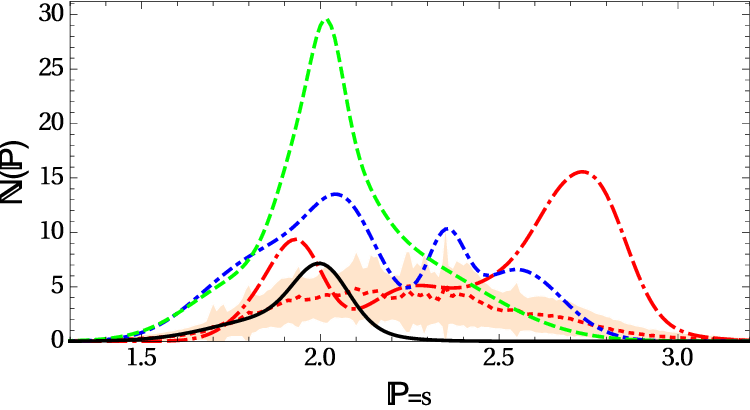}
    \includegraphics[width=0.492\textwidth]{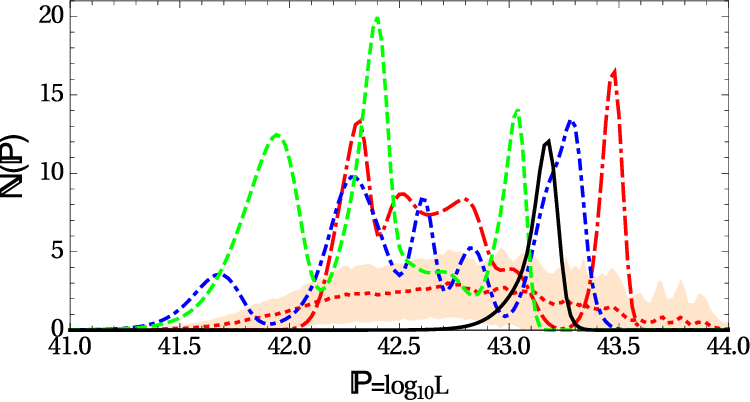}\\
    \includegraphics[width=0.492\textwidth]{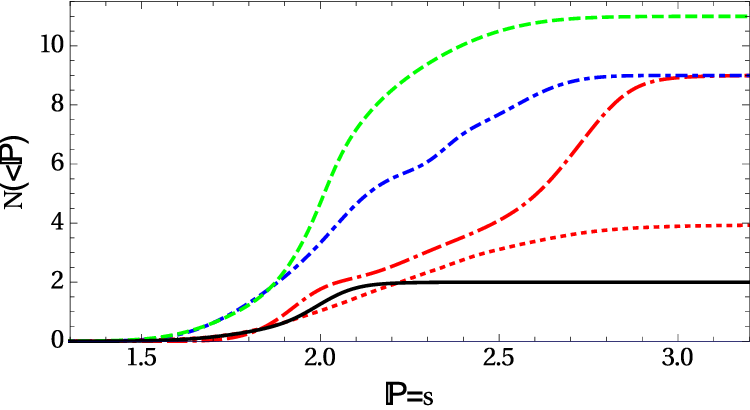}
    \includegraphics[width=0.492\textwidth]{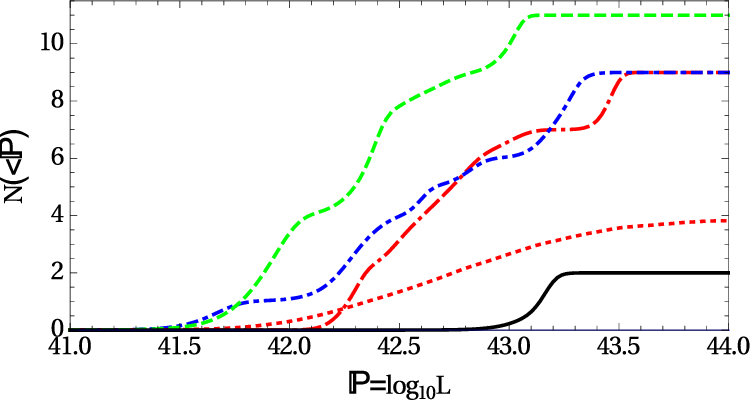}\\
    \includegraphics[width=0.492\textwidth]{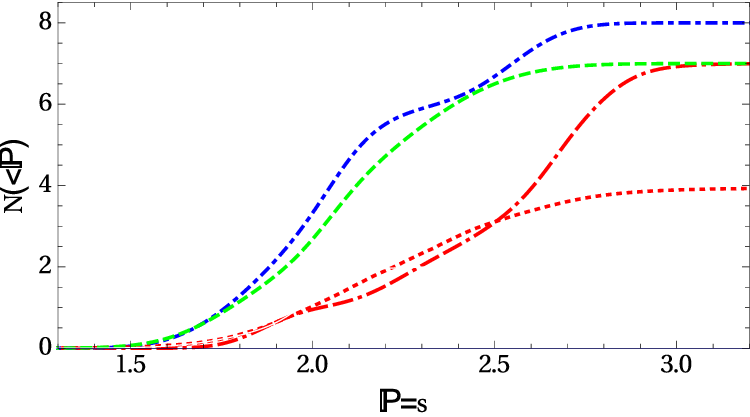}
    \includegraphics[width=0.492\textwidth]{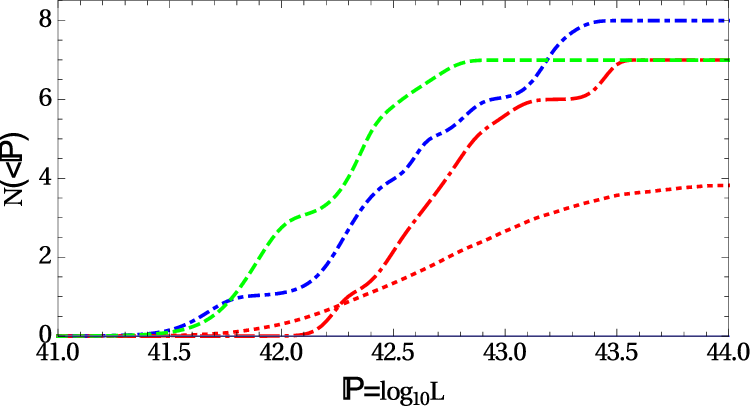}\\
    \includegraphics[width=0.492\textwidth]{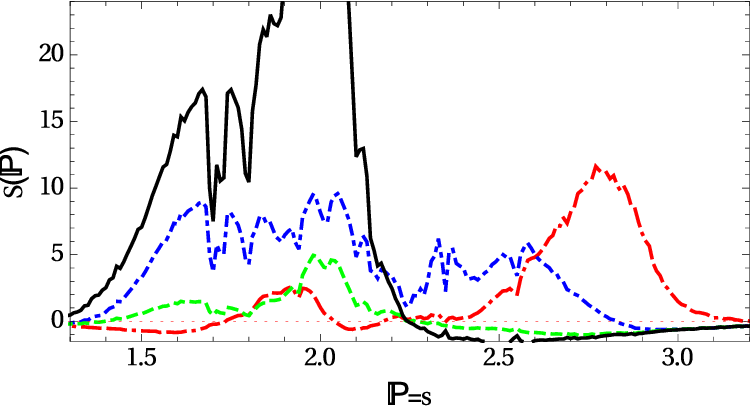}
    \includegraphics[width=0.492\textwidth]{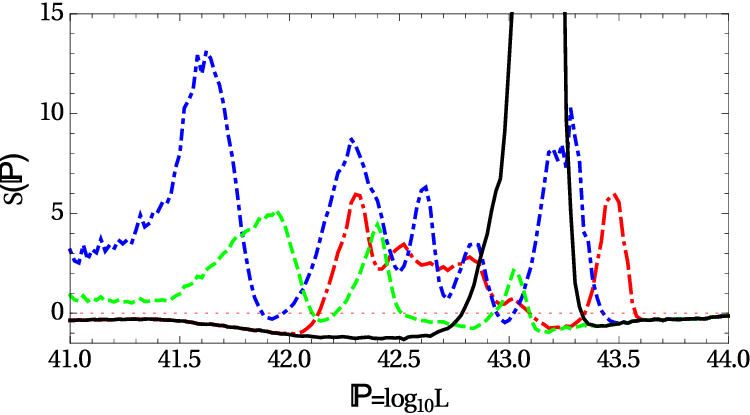}\\
    \caption{
        Same as Fig.~\ref{fig:PropB30}, but for the stacking of all 4FGL-DR4 sources around the $|b|>20\dgr$ MCXC sample (corresponding to Fig.~\ref{panel:b}).
        Here we also show the local significance (bottom row) with respect to the scaled field, which should however be considered as a lower limit because the measured field is quenched inside the virial shock.
    \label{fig:PropB20}
	}
    \vspace{-2cm}
\end{figure*}

Figure \ref{fig:PropB20} complements Fig.~\ref{fig:PropB30} by showing the properties of all (not only $>5\sigma$) sources stacked around the larger sample of 205 clusters at $|b|>20\dgr$.
Denoting $\mathbb{N}(\myP)\equiv d\NS/d\myP$ as the differential distribution of property $\myP$, for each ICM bin we compute the expected mean contribution $\mu(\mathbb{N}_{\mbox{\tiny F}})$ and standard deviation $\sigma(\mathbb{N}_{\mbox{\tiny F}})$ of the field, so the significance $S(\myP)=[\mathbb{N}-\mu(\mathbb{N}_{\mbox{\tiny F}})]/\sigma(\mathbb{N}_{\mbox{\tiny F}})$ of the local excess can be estimated.
To this end, a large ($>1500$) number of random samples of field sources is drawn, each with the same number of sources as found in the bin, and the results are then scaled to the number of sources expected in the bin based on the field source density.
The resulting field (significance) is overestimated (underestimated), because the observed field quenching is not taken into account.

The 4FGL-DR4 catalog provides, for each source, not only a pure power-law spectral fit, but also a power-law fit incorporating an exponential cutoff.
We repeat the analysis using the modified fit, and find no qualitative changes in the results.
Quantitatively, the spectral index $\mys$ hardens, for nearly all sources, by the same $\Delta \mys\simeq0.1$ factor.

\section{4GFL-DR4 sources within $\tau<1.5$}
\label{app:Sample}

The 34 sources found in the ICM, $\tau<1.5$ region of any of the 205 MCXC clusters at $|b|>20\dgr$ are listed in Table ~\ref{tab:sources}.
Of these sources, ten have optically-supported AGN classification in SIMBAD, as marked in the table.

\onecolumn

\begin{table}[t!]
\begin{threeparttable}
{\small
	\begin{longtable}{llcccclccccc}
		 \caption{
            {
            \label{tab:sources}
			 Sources in the ICM, $\tau<1.5$ regions of any of the 205 MCXC galaxy clusters in our main sample. \\
            \textbf{Columns:} (1) Cluster name; (2) MCXC name; (3) Cluster longitude $l_c$ (deg); (4) Cluster latitude $b_c$ (deg); (5) Cluster redshift $z_c$; (6) Opening angle corresponding to $R_{500}$ (deg); (7) 4FGL-DR4 source name; (8) Source longitude $l_s$ (deg); (9) Source latitude $b_s$ (deg); (10) Normalized cluster radius $\tau$ of the source ($R_{500}$ units); (11) Source power-law spectral index and its uncertainty; (12) classification in the SIMBAD database (B=BL Lac, Q=quasar, S=Seyfert galaxy; b=background object, m=cluster member, c=claim without optical support).}
			}\\
		\hline
		Cluster Name &  MCXC Name & $l_c$ & $b_c$ &  $z_c$ &  $\theta_{500}$ & 4FGL Name & $l_s$ & $b_s$ & $\tau$ & $\mys\pm\sigma(\mys)$ & type \\
		(1) & (2) & (3) & (4) & (5) & (6) & (7) & (8) & (9) & (10) & (11) & (12) \\
		\hline
		\endhead
A0194 &  J0125.6-0124 & $142.07$ & $-63.00$ & $0.02$ & $0.39$ & J0127.3-0148  & $143.22$ & $-63.25$ & $1.47$ & $2.19 \pm 0.16$ &   \\
A0400 &  J0257.6+0600 & $170.27$ & $-44.95$ & $0.02$ & $0.38$ & J0259.0+0552  & $170.75$ & $-44.83$ & $0.96$ & $2.02 \pm 0.04$ & B \\
A2029 &  J1510.9+0543 & $6.44$ & $50.53$ & $0.08$ & $0.26$ & J1510.9+0551  & $6.62$ & $50.61$ & $0.53$ & $2.01 \pm 0.16$ &   \\
A2063 &  J1523.0+0836 & $12.81$ & $49.68$ & $0.04$ & $0.35$ & J1522.5+0844  & $12.89$ & $49.87$ & $0.55$ & $2.00 \pm 0.18$ &   \\
A2147 &  J1602.3+1601 & $28.97$ & $44.54$ & $0.04$ & $0.37$ & J1603.5+1552  & $28.94$ & $44.21$ & $0.88$ & $2.33 \pm 0.17$ &   \\
A2151a &  J1604.5+1743 & $31.48$ & $44.66$ & $0.04$ & $0.29$ & J1604.7+1734  & $31.30$ & $44.57$ & $0.54$ & $2.06 \pm 0.16$ &   \\
A2589 &  J2323.8+1648 & $94.62$ & $-41.20$ & $0.04$ & $0.30$ & J2325.6+1644  & $95.10$ & $-41.45$ & $1.48$ & $1.99 \pm 0.10$ & bB \\
A3112 &  J0317.9-4414 & $252.93$ & $-56.08$ & $0.08$ & $0.22$ & J0317.8-4414  & $252.94$ & $-56.09$ & $0.07$ & $1.89 \pm 0.18$ & mQ \\
A3128 &  J0330.0-5235 & $264.80$ & $-51.12$ & $0.06$ & $0.20$ & J0331.1-5243  & $264.88$ & $-50.91$ & $1.09$ & $2.42 \pm 0.17$ &   \\
A3395 &  J0627.2-5428 & $263.24$ & $-25.19$ & $0.05$ & $0.26$ & J0625.8-5441  & $263.44$ & $-25.44$ & $1.16$ & $2.78 \pm 0.08$ & bQ \\
A3570 &  J1346.8-3752 & $314.85$ & $23.71$ & $0.04$ & $0.23$ & J1347.6-3751  & $315.02$ & $23.69$ & $0.67$ & $2.35 \pm 0.05$ & bQ \\
A3574E &  J1349.3-3018 & $317.50$ & $30.92$ & $0.02$ & $0.40$ & J1347.1-2959  & $317.05$ & $31.36$ & $1.44$ & $1.95 \pm 0.07$ & B \\
A3574W &  J1347.2-3025 & $316.95$ & $30.93$ & $0.01$ & $0.35$ & J1347.1-2959  & $317.05$ & $31.36$ & $1.23$ & $1.95 \pm 0.07$ & B \\
A3581 &  J1407.4-2700 & $323.14$ & $32.86$ & $0.02$ & $0.43$ & J1407.5-2706  & $323.13$ & $32.76$ & $0.21$ & $1.92 \pm 0.15$ &   \\
A3880 &  J2227.8-3034 & $18.00$ & $-58.51$ & $0.06$ & $0.21$ & J2227.9-3031  & $18.10$ & $-58.52$ & $0.26$ & $2.09 \pm 0.14$ & bS \\
ACO S1111 &  J2319.1-4206 & $348.45$ & $-65.94$ & $0.05$ & $0.29$ & J2319.1-4207  & $348.44$ & $-65.94$ & $0.01$ & $2.00 \pm 0.08$ & mB \\
CGCG120-014 &  J0838.1+2506 & $199.58$ & $33.73$ & $0.03$ & $0.23$ & J0837.4+2454  & $199.75$ & $33.52$ & $1.09$ & $2.72 \pm 0.19$ &   \\
MKW 11 &  J1329.5+1147 & $335.02$ & $72.24$ & $0.02$ & $0.32$ & J1328.6+1145  & $334.36$ & $72.32$ & $0.68$ & $1.74 \pm 0.13$ &   \\
NGC 0777 &  J0200.2+3126 & $139.74$ & $-29.18$ & $0.02$ & $0.38$ & J0202.7+3133  & $140.27$ & $-28.90$ & $1.46$ & $1.87 \pm 0.21$ &   \\
NGC1132 &  J0252.8-0116 & $176.44$ & $-51.08$ & $0.02$ & $0.32$ & J0253.2-0124  & $176.71$ & $-51.09$ & $0.53$ & $2.07 \pm 0.08$ & cB \\
NGC 4325 &  J1223.1+1037 & $279.58$ & $72.20$ & $0.03$ & $0.31$ & J1223.0+1100  & $279.11$ & $72.55$ & $1.24$ & $1.87 \pm 0.09$ & cB \\
NGC 5098 &  J1320.2+3308 & $78.68$ & $81.35$ & $0.04$ & $0.22$ & J1320.7+3314  & $78.70$ & $81.21$ & $0.65$ & $2.57 \pm 0.12$ &   \\
NGC 5129 &  J1324.1+1358 & $334.73$ & $74.79$ & $0.02$ & $0.28$ & J1323.9+1405  & $334.69$ & $74.92$ & $0.47$ & $1.89 \pm 0.06$ & bB \\
NGC5171 &  J1329.4+1143 & $334.86$ & $72.20$ & $0.02$ & $0.23$ & J1328.6+1145  & $334.36$ & $72.32$ & $0.85$ & $1.74 \pm 0.13$ &   \\
RXC J0340.6-0239 &  J0340.6-0239 & $189.22$ & $-42.73$ & $0.04$ & $0.26$ & J0340.5-0256  & $189.50$ & $-42.91$ & $1.05$ & $2.23 \pm 0.11$ &   \\
RXC J1257.1-1339 &  J1257.1-1339 & $305.06$ & $49.19$ & $0.02$ & $0.36$ & J1256.8-1333  & $304.95$ & $49.29$ & $0.34$ & $2.45 \pm 0.20$ &   \\
RXC J1353.4-2753 &  J1353.4-2753 & $319.30$ & $33.00$ & $0.05$ & $0.20$ & J1352.7-2742  & $319.18$ & $33.22$ & $1.21$ & $2.67 \pm 0.07$ &   \\
RXC J2018.4-4102 &  J2018.4-4102 & $359.81$ & $-33.24$ & $0.02$ & $0.33$ & J2017.5-4113  & $359.56$ & $-33.11$ & $0.75$ & $2.05 \pm 0.09$ &   \\
RXC J2104.9-5149 &  J2104.9-5149 & $346.39$ & $-41.38$ & $0.05$ & $0.22$ & J2105.2-5143  & $346.51$ & $-41.45$ & $0.51$ & $2.54 \pm 0.12$ &   \\
RX J0123.6+3315 &  J0123.6+3315 & $130.65$ & $-29.13$ & $0.02$ & $0.50$ & J0127.1+3310  & $131.47$ & $-29.10$ & $1.46$ & $2.09 \pm 0.14$ & bB \\
S0753 &  J1403.5-3359 & $319.61$ & $26.54$ & $0.01$ & $0.47$ & J1402.6-3330  & $319.57$ & $27.06$ & $1.12$ & $2.80 \pm 0.09$ &   \\
S0805 &  J1847.3-6320 & $332.25$ & $-23.60$ & $0.01$ & $0.42$ & J1848.7-6307  & $332.51$ & $-23.70$ & $0.63$ & $1.89 \pm 0.17$ &   \\
S0963 &  J2139.8-2228 & $28.12$ & $-46.53$ & $0.03$ & $0.22$ & J2139.2-2214  & $28.38$ & $-46.34$ & $1.17$ & $2.58 \pm 0.12$ &   \\
S141a &  J0113.9-3145 & $257.67$ & $-83.27$ & $0.02$ & $0.24$ & J0112.6-3158  & $260.63$ & $-83.31$ & $1.42$ & $2.01 \pm 0.15$ &   \\
VV 196 &  J0907.8+4936 & $169.27$ & $42.12$ & $0.04$ & $0.20$ & J0906.7+4950  & $168.97$ & $41.92$ & $1.47$ & $2.41 \pm 0.20$ &   \\
        \hline
   \end{longtable}
 \begin{tablenotes}
    \item[1]
    The source 4FGL J1347.1-2959 appears twice, as it is projected in the ICM outskirts of both A3574E and A3574W.
    The source is classified as BL Lac object 2MASS J13470688-2958424 \citep[\eg][]{DominguezEtAl24}.
\end{tablenotes}
}
\end{threeparttable}
\end{table}

\end{document}

%% file: Definitions.tex
\newcommand{\ie}{\emph{i.e.} }
\newcommand{\eg}{\emph{e.g.,} }

\newcommand{\be}{\begin{equation}}
\newcommand{\ee}{\end{equation}}
\newcommand{\bea}{\begin{equation*}}
\newcommand{\eea}{\end{equation*}}
\newcommand{\beqr}{\begin{eqnarray} \nonumber}
\newcommand{\eeqr}{\end{eqnarray}}
\newcommand{\beqrb}{\begin{eqnarray}}
\newcommand{\eeqrb}{\nonumber \end{eqnarray}}

\newcommand{\coma}{\mbox{ ,}}

\newcommand{\cm}{\mbox{ cm}}
\newcommand{\sr}{\mbox{ sr}}
\newcommand{\se}{\mbox{ s}}
\newcommand{\yr}{\mbox{ yr}}
\newcommand{\Gyr}{\mbox{ Gyr}}
\newcommand{\Myr}{\mbox{ Myr}}
\newcommand{\erg}{\mbox{ erg}}

\newcommand{\MHz}{\mbox{ MHz}}
\newcommand{\GHz}{\mbox{ GHz}}

\newcommand{\km}{\mbox{ km}}

\newcommand{\kpc}{\mbox{ kpc}}
\newcommand{\Mpc}{\mbox{ Mpc}}

\newcommand{\keV}{\mbox{ keV}}
\newcommand{\MeV}{\mbox{ MeV}}
\newcommand{\GeV}{\mbox{ GeV}}
\newcommand{\TeV}{\mbox{ TeV}}

\newcommand{\muG}{\mbox{ $\mu$G}}

\newcommand{\const}{\mbox{const}}

\newcommand{\gama}{$\gamma$}

\newcommand{\mynewcommand}[2]{\ifdefined #1 \else \newcommand{#1}{#2} \fi}

\mynewcommand{\apj}{ApJ}     
\mynewcommand{\apjl}{ApJL}     
\mynewcommand{\apjs}{ApJS}    
\mynewcommand{\aap}{A\&A}    
\mynewcommand{\nat}{Nature}  

\newcommand{\dgr}{^{\circ}}

\newcommand{\dgrdot}{{\overset{^\circ}{.}}}